\documentclass{article}

\usepackage{arxiv}

\usepackage[utf8]{inputenc} 
\usepackage[T1]{fontenc}    
\usepackage{hyperref}       
\usepackage{url}            
\usepackage{booktabs}       
\usepackage{amsfonts}       
\usepackage{nicefrac}       
\usepackage{microtype}      
\usepackage{lipsum}		
\RequirePackage{graphicx}
\usepackage[usenames,dvipsnames]{xcolor}
\usepackage{doi}

\RequirePackage{amsthm,amsmath,amsfonts,amssymb}
\RequirePackage[authoryear]{natbib}
\RequirePackage{mathrsfs}
\RequirePackage{algorithm}%
\RequirePackage{algorithmicx}%
\RequirePackage{algpseudocode}%
\RequirePackage{listings}%
\usepackage{bbm}

\usepackage{booktabs}
\RequirePackage{multirow}   
\usepackage{comment}
\usepackage[figuresright]{rotating}
\usepackage{rotating}
\newsavebox\CBox

\DeclareMathOperator*{\argmin}{arg\,min} 
\DeclareMathOperator*{\argmax}{arg\,max} 

\title{StabJGL: a stability approach to sparsity and similarity selection in multiple network reconstruction}

\date{June 5, 2023} 					

\author{ \href{https://orcid.org/0000-0003-2701-5686}{\includegraphics[scale=0.06]{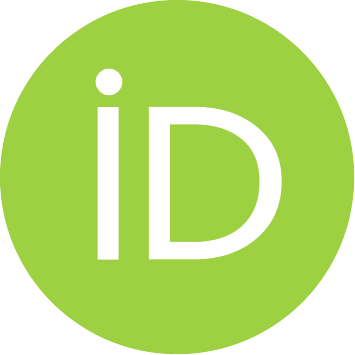}\hspace{1mm}Camilla~Lingj\ae rde}\\
	MRC Biostatistics Unit\\
	University of Cambridge\\
	\texttt{camilla.lingjaerde@mrc-bsu.cam.ac.uk} \\
	\And
	\href{https://orcid.org/0000-0003-1998-492X}{\includegraphics[scale=0.06]{orcid.pdf}\hspace{1mm}Sylvia~Richardson} \\
	MRC Biostatistics Unit\\
	University of Cambridge\\
	\texttt{sylvia.richardson@mrc-bsu.cam.ac.uk} \\
}

\renewcommand{\headeright}{}
\renewcommand{\undertitle}{}
\renewcommand{\shorttitle}{StabJGL: stable sparsity and similarity selection for multiple networks}

\hypersetup{
pdftitle={StabJGL: a stability approach to sparsity and similarity selection in multiple network reconstruction},
pdfsubject={stat.ME},
pdfauthor={Camilla~Lingj\ae rde, Sylvia~Richardson},
pdfkeywords={High-dimensional inference, Network models, Joint graphical model, Joint graphical lasso, Gaussian graphical model, Genomics, Gene networks, Protein–protein interaction networks, Integrative analysis},
}

\begin{document}
\maketitle

\begin{abstract}
 In recent years, network models have gained prominence for their ability to capture complex associations. In statistical omics, networks can be used to model and study the functional relationships between genes, proteins, and other types of omics data. If a Gaussian graphical model is assumed, a gene association network can be determined from the non-zero entries of the inverse covariance matrix of the data. Due to the high-dimensional nature of such problems, integrative methods that leverage similarities between multiple graphical structures have become increasingly popular. The joint graphical lasso is a powerful tool for this purpose, however, the current AIC-based selection criterion used to tune the network sparsities and similarities leads to poor performance in high-dimensional settings. 
 We propose stabJGL, which equips the joint graphical lasso with a stable and accurate penalty parameter selection approach that combines the notion of model stability with likelihood-based similarity selection. The resulting method makes the powerful joint graphical lasso available for use in omics settings, and outperforms the standard joint graphical lasso, as well as state-of-the-art joint methods, in terms of all performance measures we consider. Applying stabJGL to proteomic data from a pan-cancer study, we demonstrate the potential for novel discoveries the method brings. A user-friendly R package for stabJGL with tutorials is available on Github: \url{https://github.com/Camiling/stabJGL}.
\end{abstract}


\section{Introduction}
\label{sec:intro}

Network models have in recent years gained great popularity in many areas. In statistical omics, networks can be used to decode aspects of unknown structures, and hence study the relationships between genes, proteins, and other types of omics data. In health data sciences, rich data sets are more and more frequently encountered, enabling the development of models integrating a variety of biological resources. In the high-dimensional setting commonly found in omics, sharing information between data sources with shared structures -- which could be different tissues, conditions, patient subgroups, or different omics types -- can give a valuable increase in statistical power while elucidating shared biological function. A key question is how to combine the different data sources into a single model. 

If a Gaussian graphical model is assumed, a conditional (in)dependence network can be estimated by determining the non-zero entries of the inverse covariance (precision) matrix of the data. With its good performance in numerical studies, the \emph{graphical lasso} of \cite{friedman2008} is a state-of-the-art method for precision matrix estimation in the setting of Gaussian graphical models. The method combines $L_1$ regularization with maximum likelihood estimation. Other notable methods include the neighborhood selection approach of \cite{Meinshausen06} and the graphical SCAD \citep{fan2009}. Notable Bayesian methods include the Bayesian graphical lasso \citep{wang2012bayesian}, Bayesian spike-and-slab approaches \citep{wang2015spike} and the graphical horseshoe \citep{li2019graphical}. If multiple related data sets are available, there are several ways to leverage common network structures. If focusing on one data type's network structure, data from other types can enhance inference via weighted graphical lasso methods \citep{Li15,lingjaerde2021tailored}. However, to compare network structures across data sets, such as patient subgroups, a joint approach that leverages common information while preserving the differences can increase statistical power and provide interpretable insight.

In the area of multiple Gaussian graphical models, existing methods include the group extension of the graphical lasso to multiple networks of \cite{guo2011joint}, the Bayesian spike-and-slab joint graphical lasso \citep{li2019bayesian} and the Markov random field approach of \cite{peterson2015bayesian}. The widely used joint graphical lasso (JGL) of \cite{danaher2014} extends the graphical lasso to a multiple network setting and provides a powerful tool for inferring graphs with common traits. It employs two different penalty functions -- group (GGL) and fused (FGL) -- with the latter recommended for most applications. From this point forward, any mention of the joint graphical lasso will imply the fused version, unless otherwise specified. The method needs tuning of two regularization parameters for controlling (i) the number of non-zero effects, and (ii) the similarity between networks, respectively. However, the default parameter selection routine based on the AIC \citep{akaike1973} often results in severe over-selection in high-dimensional data, potentially impacting performance negatively \citep{liu2010stability, foygel2010extended}. In such settings, selection approaches based on model stability have demonstrated competitive performance \citep{liu2010stability, angelini2022jewel}.

We propose a stable and accurate penalty parameter selection method for the joint graphical lasso, combining the model stability principle of \cite{liu2010stability} with likelihood-based selection for high-dimensional data \citep{foygel2010extended}. The resulting method inherits the powerful traits of the joint graphical lasso while mitigating the risk of severe under- or over-selection of edges in high-dimensional settings. We provide an R package, \texttt{stabJGL} (stable sparsity and similarity selection for the joint graphical lasso), which implements the method. 

The paper is organized as follows. In Section \ref{sec:methods}, we first describe the Gaussian graphical model framework and the penalized log-likelihood problem we aim to solve. We then describe our proposed algorithm. In Section \ref{sec:results}, we demonstrate the performance of our proposed method on simulated data and apply it proteomic data from a pan-cancer study of hormonally responsive cancers. Finally, we highlight possible extensions in Section \ref{sec:discussion}. 

\section{Materials and methods}
\label{sec:methods}
\subsection{Gaussian graphical models}

In a gene network model, genes are represented by \emph{nodes} and associations between them are represented by \emph{edges}. Given measurable molecular units each corresponding to one gene (e.g., the encoded protein or mRNA), a network, or graph, can be constructed from their observed values. Consider $n$ observed values of the multivariate random vector $\boldsymbol{x} = (X_1, \ldots, X_{p})^T$ of node attributes, with each entry corresponding to one of $p$ nodes. If we assume multivariate Gaussian node attributes, with an $n\times p$ observation matrix $\boldsymbol{X}$ with i.i.d. rows $\boldsymbol{x_1}, \ldots, \boldsymbol{x_n} \sim \mathcal{N}(\boldsymbol{0},\boldsymbol{\Sigma})$, a \emph{partial correlation network} can be determined by estimating the inverse covariance matrix, or precision matrix, $\boldsymbol{\Theta}=\boldsymbol{\Sigma}^{-1}$. Specifically, the partial correlation between nodes $i$ and $j$, conditional upon the rest of the graph, is given by 
\begin{equation}
\label{eq:conditionalcor}
\rho_{ij\mid V\backslash \{i,j\} } = - \frac{\theta_{ij}}{\sqrt{\theta_{ii}\theta_{jj}}}, \nonumber
\end{equation}
where the $\theta_{ij}$'s are the entries of $\boldsymbol{\Theta}$ and $V$ the set of all node pairs \citep{lauritzen1996graphical}. The partial correlations coincide with the conditional correlations in the Gaussian setting. Because correlation (resp. partial correlation) equal to zero is equivalent to independence (resp. conditional independence) for Gaussian variables, a conditional independence graph can thus be constructed by determining the non-zero entries of the precision matrix. To ensure invertibility, the precision matrix also required to be positive definite, $\boldsymbol{\Theta} \succ 0$. %

In high-dimensional settings, the sample covariance matrix $\boldsymbol{S} = \frac{1}{n-1}\boldsymbol{X}^T \boldsymbol{X}$ is rarely of full rank and thus its inverse cannot be estimated directly. It is common to assume sparse network, meaning the number of edges in the edge set $E$ is small relative to the number of potential edges in the graph (i.e., the sparsity measure $2\vert E\vert/(p^2-p)$ is small). Penalized methods such as the graphical lasso \citep{friedman2008} are well established for sparse Gaussian graphical model estimation. In the case of there being multiple (related) data sets available, such as from different tissue types, rather than estimating each network separately  much statistical power could be gained by sharing information across networks through a joint approach.

\subsection{Penalized log-likelihood problem}

Assume a network inference problem with $K$ groups. We let $\{\boldsymbol{\Theta}\} = (\boldsymbol{\Theta}^{(1)},\ldots, \boldsymbol{\Theta}^{(K)})$ be the set of their (unknown) precision matrices, and assume that the set of $\sum_{k=1}^K n_k$ observations are independent. We aim to solve the penalized log-likelihood problem \citep{danaher2014}
\begin{align}
\label{eq:penalizedloglikJGL}
 \{\widehat{\boldsymbol{\Theta}} \}&= \argmax_{\{\boldsymbol{\Theta} \succ 0\}} \Big \{ \sum\limits_{k=1}^{K} n_k [\log (\det (\boldsymbol{\Theta}^{(k)} )) - \text{tr} (\boldsymbol{S}^{(k)}  \boldsymbol{\Theta}^{(k)} )] \\ 
 &- \text{P}(\{ \boldsymbol{\Theta } \})\Big \} \nonumber
\end{align}
where $\boldsymbol{S}^{(k)}$ is the sample covariance matrix of group $k$ and $\text{P}(\cdot)$ is a penalty function. In (\ref{eq:penalizedloglikJGL}), $\det(\cdot)$ denotes the determinant and $\text{tr}(\cdot)$ denotes the trace. The joint graphical lasso employs the fused penalty function
\begin{align}
    \label{eq:FGL}
    \text{P}(\{ \boldsymbol{\Theta } \}) = \lambda_1\sum\limits_{k=1}^K\sum\limits_{i\neq j}\text{abs}(\theta_{ij}^{(k)}) + \lambda_2 \sum\limits_{k<k'}\| \boldsymbol{\Theta}^{(k)} - \boldsymbol{\Theta}^{(k')}\|_1
\end{align}
where $\lambda_1$ and $\lambda_2$ are positive penalty parameters, $\text{abs}(\cdot)$ denotes the absolute value function and $\| \cdot \|_1$ denotes the $\text{L}_1$ penalty. This penalty applies $\text{L}_1$ penalties to each off-diagonal element of the $K$ precision matrices as well as to the differences between corresponding elements of each pair of precision matrices. As for the graphical lasso, the parameter $\lambda_1$ controls the sparsity. The similarity parameter $\lambda_2$ controls the degree to which the $K$ precision matrices are forced towards each other, encouraging not only similar network structures but also similar precision matrix entries. The current penalty parameter selection approach for $\lambda_1$ and $\lambda_2$ is based on the AIC \citep{danaher2014}. While suitable for determining network similarities, likelihood-based selection criteria can lead to severe under- or over-selection and thus poor performance in high-dimensional settings \citep{liu2010stability}.


\subsection{The stabJGL algorithm}

\begin{figure}[t]
    \centering
    \includegraphics[width=0.7\textwidth]{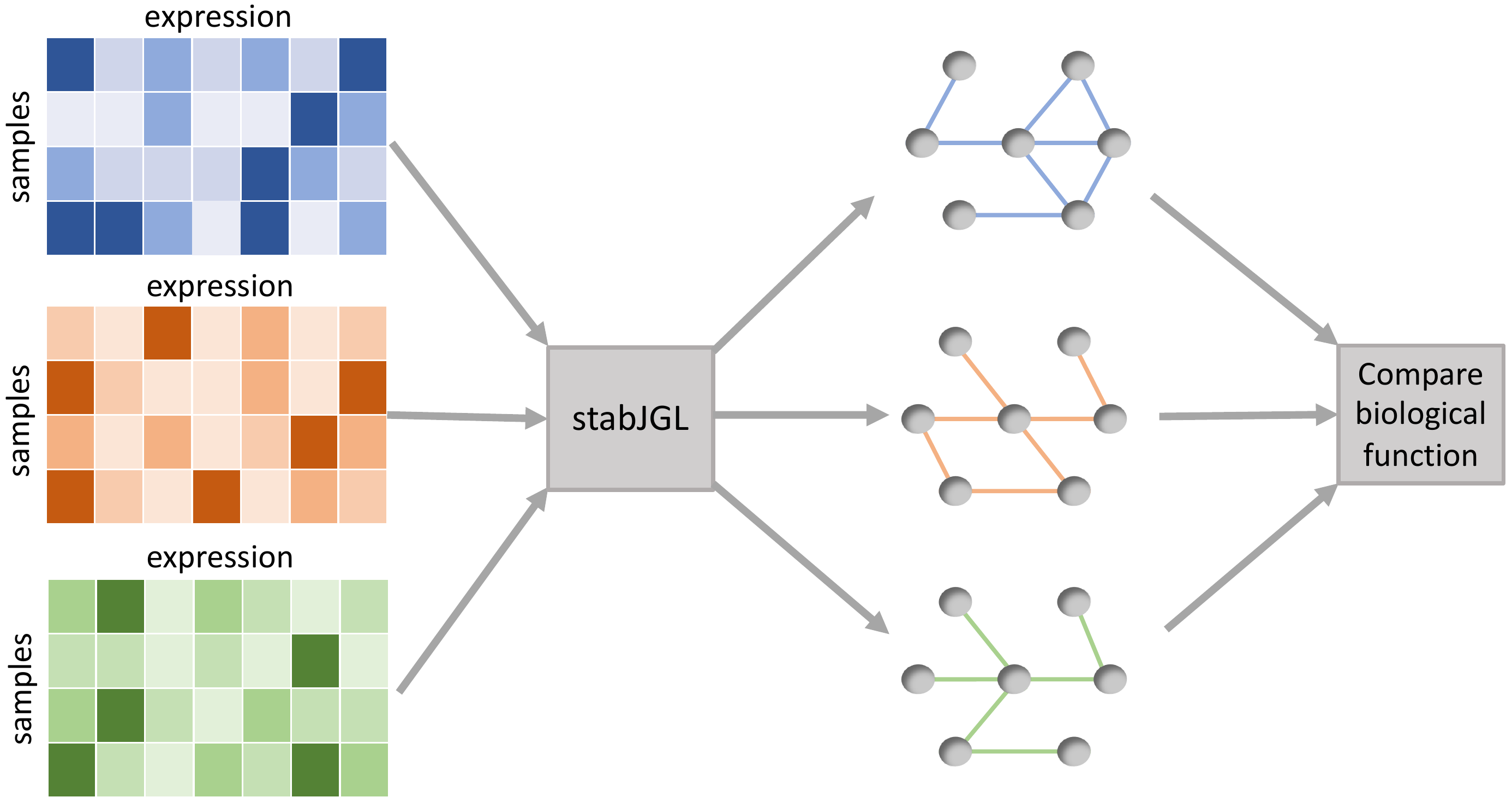}
    \caption{The workflow of stabJGL, where the network structures of different data types or conditions are jointly estimated and can then be compared.}
    \label{fig:netflow}
\end{figure}

To improve the performance of the joint graphical lasso with the fused penalty for omics applications and other high-dimensional problems, we propose the stabJGL algorithm for stable sparsity and similarity selection in multiple network reconstruction. Below we outline the algorithm, where we first select the sparsity parameter $\lambda_1$ in the fused penalty (\ref{eq:FGL}) based on the notion of model stability, and then the similarity parameter $\lambda_2$ based on model likelihood. StabJGL jointly estimates multiple networks by leveraging their common information, and gives a basis for deeper exploration of their differences, as shown in Figure \ref{fig:netflow}.

\subsubsection{Selecting $\lambda_1$}

We select $\lambda_1$ by extending the framework introduced by \cite{liu2010stability} in their Stability Approach to regularization Criterion (StARS) to a multiple network setting. The aim is to select the least amount of penalization that makes graphs sparse as well as reproducible under random subsampling. This is done by drawing many random subsamples from each of the $K$ data types and using them to construct joint graphical lasso graphs over a range of $\lambda_1$ values. The smallest parameter value for which a given graph estimation variability measure does not surpass a specified threshold is then selected. We use a measure of edge assignment instability across subsamples to quantify the variability.

Specifically, we consider a grid of $\lambda_1$ values in a suitable interval, i.e., $(0,1]$ and keep the similarity parameter $\lambda_2$ fixed to some small value such as $0.01$ in the first instance. For $\eta=1,\ldots,N_{\text{sample}}$, we draw a random subsample from each group $k$'s set of $n_k$ observations without replacement, each of size $b_k < n_k$. \cite{liu2010stability} show that in a single network setting, $b_k = \lfloor 10 \sqrt{n_k} \rfloor$ maintains theoretical properties for containing the true graph with high probability as well as high empirical performance, and this is the value we use. For each value of $\lambda_1$ to consider, we next construct the corresponding set of joint graphical lasso graphs $\{G_{(k)}^\eta(\lambda_1)\}_{k=1}^{K}$ from these $K$ sets of subsamples, using the fused penalty (\ref{eq:FGL}). 

The following is then done for each value of $\lambda_1$ we consider. For each group $k=1,\ldots,K$ and all possible node pairs $(i, j)$ we estimate the probability of an edge between the nodes over the $N_{\text{sample}}$ inferred sets of graphs 
\begin{align}
    \widehat{\psi}_{ij}^{(k)}(\lambda_1) = \frac{1}{N_{\text{sample}}}\sum_{\eta=1}^{N_{\text{sample}}}\mathbbm{1}\left[(i,j) \in G^\eta_{(k)}(\lambda_1)\right],
\end{align}
where $\mathbbm{1}\left[\cdot \right]$ is the indicator function. Using this estimated probability, we find
\begin{align}
\widehat{\xi}_{ij}^{(k)}(\lambda_1) =2 \widehat{\psi}_{ij}^{(k)}(\lambda_1)(1-\widehat{\psi}_{ij}^{(k)}(\lambda_1)),
\end{align}
which is an estimate of two times the variance of the Bernoulli indicator of the edge $(i,j)$ in group $k$. It lies in $[0,0.5]$ and can be regarded as an estimate of the fraction of times two inferred graphs for group $k$ found with the joint graphical lasso with the given $\lambda_1$ value will disagree on the presence of the edge $(i,j)$. Due to the $L_1$ penalty in (\ref{eq:FGL}), the number of inferred edges will decrease as $\lambda_1$ is increased. For a given $\lambda_1$, $\widehat{\xi}_{ij}^{(k)}(\lambda_1)$ can be regarded as a measure of the variability of the edge $(i,j)$ in group $k$ across subsamples, and the total variability of graph $k$ can be measured by averaging over all edges, yielding the estimate 
\begin{align}
\widehat{D}_{(k)}(\lambda_1) = \frac{1}{\binom{p}{2}} \sum_{i<j} \widehat{\xi}_{ij}^{(k)}(\lambda_1).
\end{align}

For each value of $\lambda_1$, the total variability of the whole set of graphs found by the joint graphical lasso is then found by averaging the variability over all $K$ networks
\begin{align}
    \widehat{D}(\lambda_1) = \frac{1}{K} \sum_{k=1}^{K} \widehat{D}_{(k)}(\lambda_1).
\end{align}

For sufficiently large $\lambda_1$, all edges are excluded from the model and so the variability $\widehat{D}(\lambda_1)$ will be $0$. The variability will in general increase as the penalty $\lambda_1$ decreases, however, for small enough $\lambda_1$ the graphs will become so dense that the variability starts to decrease again. As sparse network inference is the aim, we therefore monotonize the variability function by letting $\bar{D}(\lambda_1) = \sup_{0\leq t \leq \lambda_1}\widehat{D}(t)$. Finally, for a given variability threshold $\beta_1$, the optimal penalty is chosen to be $\widehat{\lambda}_{1} = \sup \{ \lambda_1 :  \bar{D}(\lambda_1) \leq \beta_1 \}$. As opposed to $\lambda_1$, $\beta_1$ is an interpretable quantity and we propose a default threshold of $\beta_1=0.1$ as suggested by \citeauthor{liu2010stability} for the original StARS algorithm, which reflects an acceptance of $10\%$ variability in the edge assignments.

\subsubsection{Selecting $\lambda_2$}

After $\lambda_1$ has been selected, we select $\lambda_2$ with a multiple-network version of the extended BIC (eBIC or $\text{BIC}_{\gamma}$) of \cite{foygel2010extended}. The eBIC is an extension of the Bayesian Information Criterion of \cite{schwarz1978estimating}, where the prior is reformulated to account for high-dimensional graphical settings. We propose an adaptation the eBIC to a multiple-network setting,
\begin{align}
    \text{BIC}_{\gamma}(\lambda_1,\lambda_2) &= \sum_{k=1}^K \Big[ n_k \text{tr}(\boldsymbol{S}^{(k)}\widehat{\boldsymbol{\Theta}}^{(k)}_{\lambda_1,\lambda_2})  
    - n_k \log(\det(\widehat{\boldsymbol{\Theta}}^{(k)}_{\lambda_1,\lambda_2})) \nonumber \\ &+ \vert E_k \vert \log{n_k} + 4 \vert E_k \vert \gamma \log{p} \Big],
    \label{eq:ebicAd}
\end{align}
where $\widehat{\boldsymbol{\Theta}}^{(k)}_{\lambda_1,\lambda_2}$ is the estimated precision matrix of network $k$ obtained with the penalty parameters $\lambda_1$ and $\lambda_2$, and $\vert E_k\vert$ is the size of the corresponding edge set. A grid of $\lambda_2$ values is considered, with $\lambda_1$ fixed to the value selected in the previous step. The value of $\lambda_2$ that minimizes (\ref{eq:ebicAd}) is selected. 
Like for the standard eBIC, the additional edge penalty parameter $\gamma\in [0,1]$ must be chosen. However, since we are using the eBIC for similarity selection rather than sparsity selection, the choice of $\gamma$ is not as important because we are comparing graphs with the same value of $\lambda_1$ and hence similar levels of sparsity. We typically use $\gamma=0$, which corresponds to the ordinary BIC, for most applications. Our implementation includes the eBIC generalization to give the user the option of additional penalization in extremely high-dimensional cases.

\subsubsection{Algorithm}

The full stabJGL algorithm is given in Algorithm \ref{algo1}. $\text{JGL}(\cdot)$ indicates that the joint graphical lasso function with the fused penalty is applied. The output of the JGL function can either be a set of graphs, a set of precision matrices or an edge set, depending on what is required Algorithm \ref{algo1}. 

\begin{algorithm}[!t]
\caption{The stabJGL algorithm}\label{algo1}
\begin{algorithmic}[1]
\Require $n_k \times p$ data matrix $\boldsymbol{X}^{(k)}$ for $k=1,\ldots,K$
\State $\Lambda_1 \leftarrow \{ 0.01,0.02,\ldots,1\}$, $\Lambda_2 \leftarrow \{ 0,0.01,\ldots,0.1\}$
\State $\lambda_2^{(\text{init})} \leftarrow  0.01$
\State $\beta_1 \leftarrow  0.1$
\State $N_{\text{sample}} \leftarrow 20$
\State $\gamma \leftarrow 0$
\State $b_k \leftarrow  \lfloor 10\sqrt{n_k} \rfloor$ for $k=1,\ldots,K$
\State $\boldsymbol{S}^{(k)} \leftarrow  \frac{1}{n_k-1}{\boldsymbol{X}^{(k)}}^T {\boldsymbol{X}^{(k)}}$ for $k=1,\ldots,K$
\For{$\lambda_1$ in $\Lambda_1$}
    \For{$\eta=1$ to $N_{\text{sample}}$}
        \For{$k=1$ to $K$}
            \State Sample $b_k$  indices $I_k \subset \{1,\ldots,n_k \}$
            \State $\boldsymbol{X}^{(k)}_{\text{sample}} \leftarrow \boldsymbol{X}^{(k)}[I_k,]$
        \EndFor
        \State $\{G_{(k)}^\eta(\lambda_1)\}_{k=1}^{K} \leftarrow \text{JGL}\left(\{ \boldsymbol{X}^{(k)}_{\text{sample}}\}_{k=1}^{K} \mid \lambda_1, \lambda_2^{(\text{init})} \right)$
    \EndFor
    \For{$k=1$ to $K$}
        \For{$j=1$ to $p$}
            \For{$i=1$ to $j-1$}
                \State $\widehat{\psi}_{ij}^{(k)}(\lambda_1) \leftarrow \frac{1}{N_{\text{sample}}}\sum_{\eta=1}^{N_{\text{sample}}}\mathbbm{1}\left[(i,j) \in G^\eta_{(k)}(\lambda_1)\right]$
                \State $\widehat{\xi}_{ij}^{(k)}(\lambda_1) \leftarrow 2 \widehat{\psi}_{ij}^{(k)}(\lambda_1)(1-\widehat{\psi}_{ij}^{(k)}(\lambda_1))$
            \EndFor
        \EndFor
        \State $\widehat{D}_{(k)}(\lambda_1) \leftarrow \frac{1}{\binom{p}{2}} \sum_{i<j} \widehat{\xi}_{ij}^{(k)}(\lambda_1)$
    \EndFor
    \State $\widehat{D}(\lambda_1) \leftarrow \frac{1}{K} \sum_{k=1}^{K} \widehat{D}_{(k)}(\lambda_1)$
    \State $\bar{D}(\lambda_1) \leftarrow \sup_{0\leq t \leq \lambda_1}\widehat{D}(t)$
\EndFor
\State $\widehat{\lambda}_{1}  \leftarrow  \sup \{ \lambda_1 \in \Lambda_1:  \bar{D}(\lambda_1) \leq \beta_1 \}$
\For{$\lambda_2$ in $\Lambda_2$}
    \State $\{\widehat{\boldsymbol{\Theta}}^{(k)}_{\widehat{\lambda}_{1}\lambda_2}, E_k \}_{k=1}^{K} \leftarrow \text{JGL}\left(\{ \boldsymbol{X}^{(k)}\}_{k=1}^{K} \mid \widehat{\lambda}_{1}, \lambda_2 \right)$
    \State $\text{BIC}_{\gamma}(\widehat{\lambda}_{1} ,\lambda_2) \leftarrow  \sum_{k=1}^K \Big[ n_k \text{tr}(\boldsymbol{S}^{(k)}\widehat{\boldsymbol{\Theta}}^{(k)}_{\widehat{\lambda}_{1}, \lambda_2}) - n_k \log(\det(\widehat{\boldsymbol{\Theta}}^{(k)}_{\widehat{\lambda}_{1},\lambda_2}))   + \vert E_k \vert \log{n_k} + 4 \vert E_k \vert \gamma \log{p} \Big]$
\EndFor
\State  $\widehat{\lambda}_{2} \leftarrow \argmin_{ \lambda_2 \in \Lambda_2} \text{BIC}_{\gamma}(\widehat{\lambda}_{1} ,\lambda_2)$
\State $\{\widehat{\boldsymbol{\Theta}}_{\text{stabJGL}}^{(k)}\}_{k=1}^{K} \leftarrow \text{JGL}\left(\{ \boldsymbol{X}^{(k)}\}_{k=1}^{K} \mid \widehat{\lambda}_{1}, \widehat{\lambda}_2 \right)$
\end{algorithmic}
\end{algorithm}

\subsubsection{Implementation details}

StabJGL is implemented in R, and available as an R package at \url{https://github.com/Camiling/stabJGL}. The subsampling routine is implemented so it can be done in parallel.
The joint graphical lasso fittings are done as in \cite{danaher2014}, using an ADMM (Alternating Direction Method of Multipliers) algorithm \citep{boyd2011} for general penalty functions to solve the penalized log-likelihood problem (\ref{eq:penalizedloglikJGL}), By default, $20$ subsamples are used and we evaluate $20$ values each of $\lambda_1\in [0.01,1]$ and $\lambda_2 \in [0,0.1]$. As in StARS, we use a subsample size of $\lfloor 10\sqrt{n_k} \rfloor$ for group $k=1,\ldots,K$ \citep{liu2010stability}. The additional penalty parameter $\gamma$ in the eBIC for similarity selection is set to $0$ by default, corresponding to the standard BIC. We found this value to be suitable in most applications but leave the option to increase the penalization.
We employ a default variability threshold of $\beta_1=0.1$.

\section{Results}
\label{sec:results}

\subsection{Simulated data}
\label{subsec:simdata}

We first assess the performance of stabJGL on simulated data. We compare the network reconstruction ability of stabJGL to that of state-of-the-art methods, including the joint graphical lasso with the fused penalty (FGL) and group penalty (GGL) with penalty parameters selected with the default AIC-based criterion \citep{danaher2014}. To assess the performance of another selection criterion specifically designed for high-dimensional graph selection, we also consider FGL with penalty parameters tuned by the extended BIC for multiple graphs (\ref{eq:ebicAd}) with a moderate value of $\gamma=0.2$ \citep{foygel2010extended}. 
We further include the Bayesian spike-and-slab joint graphical lasso (SSJGL) of \cite{li2019bayesian}, as well as the graphical lasso (Glasso) of \cite{friedman2008} tuned by StARS \citep{liu2010stability}. The latter estimates each network separately. We generate data that closely resembles our omics application of interest, featuring partial correlations between $0.1$ and $0.2$ in absolute value, while also exhibiting the \emph{scale-free} property - a typical assumption for omics data where the \emph{degree distribution} (i.e., the distribution of the number of edges that are connected to the nodes) adheres to a power-law distribution \citep{chen2004content}. In the main simulation scenario, we simulate $K=3$ networks with $p=100$ nodes, manipulating the degree of similarity in their ``true'' graphical structures to assess the performance of the method over a wide range of scenarios. We maintain a sparsity of $0.02$ across all networks and generate data sets from the corresponding multivariate Gaussian distributions with $n_1=150$, $n_2=200$ and $n_3=300$ observations. We then apply different network reconstruction techniques to determine the networks from the data. For FGL and GGL, the two penalty parameters are chosen in a sequential fashion with the default AIC-based criterion proposed by \cite{danaher2014}, with $20$ values of $\lambda_1\in[0.01,1]$ and $\lambda_2\in[0,0.1]$ respectively being evaluated. We consider the eBIC criterion on the same grid of values for FGL. We consider the same set of $\lambda_1$ and $\lambda_2$ values in the stabJGL algorithm and let $\gamma=0$ in the eBIC criterion for similarity selection. For stabJGL and the graphical lasso tuned by StARS, we use a variability threshold of $0.1$ and use $20$ subsamples. For the Bayesian spike-and-slab joint graphical lasso all parameter specifications are as suggested by \cite{li2019bayesian}. In addition to the above setup, we consider additional settings with $K\in\{2, 4\}$ graphs and $p=100$ nodes. We only show a summarizing plot of these additional results, but the full tables for these simulations, as well as from additional scenarios with other values of $K$ and $p$, are given in the Supplement. We also investigate the effect of the variability threshold $\beta_1$ in stabJGL on the results in a setting with $p=100$ nodes and $K=2$ networks. Finally, to compare the scalability of the respective methods we consider the time needed to infer networks for various $p$ and $K$. Further details and code for the simulation study can be found at \url{https://github.com/Camiling/stabJGL_simulations}.

Estimation accuracy is assessed with the \emph{precision} (positive predictive value), and the \emph{recall} (sensitivity). The precision gives the fraction of predicted edges that were correct, while the recall is the fraction of edges in the true graph that were identified by the inference. Because the sparsity of estimated networks will vary between methods, the precision-recall trade-off should be taken into consideration. In general, the recall will increase with the number of selected edges while the precision will decrease. Since sparsity selection is a main feature of our proposed method, we do not consider threshold-free comparison metrics such as the AUC. We therefore put emphasis on the following characteristics in our comparative simulation study; (i) suitable sparsity level selection, (ii) utilization of common information at any level of network similarity, i.e., inference improves with increased network similarity, and (iii) a suitable precision-recall trade-off that overly favours either measure. 

\subsection{Simulation results}

The results are summarized in Table \ref{table:simulation}. First, we observe that the fused and group joint graphical lasso with the default AIC-based penalty parameter selection strongly over-select edges in all cases. This leads to high recall, but very low precision. Second, they do not appear to sufficiently utilize network similarities; the performance of the two methods, particularly GGL, differs little between completely unrelated and identical networks. Notably, in all cases the selected value of $\lambda_2$ is smaller for FGL and GGL tuned by AIC than it is for stabJGL. Consequently, similarity is not sufficiently encouraged even in settings where the networks are identical. The AIC criterion does not seem to provide sufficient penalization to encourage suitable sparsity and similarity. On the other hand, we observe that the alternative eBIC criterion gives extremely sparse FGL estimates, resulting in high precision but very low recall. In half of the cases, it selects an empty graph, i.e., no edges. Although the extended BIC is developed specifically for graphical model selection, likelihood-based criteria for sparsity selection tend to perform poorly in high-dimensional settings and risk both severe under- and over-selection \citep{foygel2010extended}. This issue is avoided in the stabJGL algorithm as the eBIC only is used to select similarity and not sparsity.

The Bayesian spike-and-slab joint graphical lasso tends to select very few edges, leading to high precision but low recall. Its performance deteriorates drastically as the network differences increase, leading to extremely low recall. This implies a lack of flexibility to adapt to varying network similarity levels, as has previously been observed \citep{lingjaerde2022scalable}. Out of all the joint methods, stabJGL gives the most accurate sparsity estimate. This ensures that we neither get very low precision like FGL and GGL tuned by AIC, nor very low recall like SSJGL and FGL tuned by eBIC. StabJGL also appears to adapt well to the similarity between networks, with the prediction accuracy increasing with the number of shared edges. As a result, the method either outperforms the graphical lasso tuned by StARS for highly similar networks or performs comparably to it for unrelated networks. The similar performance for unrelated networks can be explained by the fact that the sparsity controlling penalty parameter of both methods are tuned with a stability-based approach. The results suggest that stabJGL can be used agnostically in settings where there is no prior knowledge about the level of network similarity and does not run any risk of decreased accuracy should the networks have nothing in common. 

\begin{table*}[t]
\caption{Performance of the different graph reconstruction methods in simulations, reconstructing graphs with $p=100$ nodes from $K=3$ networks with various similarity of the true graph structures. The methods included are Glasso, FGL and GGL tuned by AIC, FGL tuned by eBIC, SSJGL and stabJGL. The similarity (percentage of edges that are in common) of the graphs is shown. The results are averaged over $N=100$ simulations and shows the sparsity, precision, and recall of each of the $K=3$ estimated graphs. The corresponding standard deviations are shown in parentheses. The graphs are reconstructed from $n_1=150$, $n_2=200$ and $n_3=300$ observations. All graphs have sparsity $0.02$. 
The average selected values of the penalty parameters $\lambda_1$ and $\lambda_2$ for the relevant methods is shown as well.  \label{table:simulation}}
\hspace{-1cm}
\renewcommand{\arraystretch}{1.4}
\resizebox{1.1\textwidth}{!}{%
\begin{tabular}{r l r r l @{\hskip 0.1cm}l l l l @{\hskip 0.1cm}l l l l@{\hskip 0.1cm}l l l}
\toprule
&&&&& \multicolumn{3}{@{}c@{}}{$n_1=150$}&&\multicolumn{3}{@{}c@{}}{$n_2=200$}&&\multicolumn{3}{@{}c@{}}{$n_3=300$} \\
\cline{6-8}\cline{10-12} \cline{14-16}%
Similarity & Method & $\lambda_1$& $\lambda_2$ &&Sparsity& Precision & Recall &&Sparsity& Precision & Recall&&Sparsity& Precision & Recall\\
\midrule
100 $\%$ & Glasso &  0.208  & -  && 0.026 (0.007)  & 0.41 (0.08)  & 0.51 (0.06)  && 0.018 (0.004)  & 0.57 (0.08)  & 0.51 (0.05)  && 0.014 (0.002)  & 0.76 (0.07)  & 0.54 (0.05)  \\ 
  & FGL  &  0.114  &  0.021 && 0.087 (0.028)  & 0.22 (0.08)  & 0.88 (0.04)  && 0.064 (0.020)  & 0.31 (0.09)  & 0.90 (0.03)  && 0.041 (0.010)  & 0.46 (0.09)  & 0.91 (0.03)  \\ 
  & FGL (eBIC)  &  0.365  &  0.021 && 0.004 (0.004)  & 0.97 (0.06)  & 0.20 (0.20)  && 0.004 (0.004)  & 0.99 (0.03)  & 0.20 (0.20)  && 0.004 (0.004)  & 0.99 (0.01)  & 0.20 (0.20)  \\ 
  & GGL  &  0.114  &  0.007 && 0.152 (0.024)  & 0.11 (0.03)  & 0.81 (0.04)  && 0.114 (0.019)  & 0.15 (0.04)  & 0.84 (0.04)  && 0.069 (0.013)  & 0.27 (0.06)  & 0.87 (0.03)  \\ 
  & SSJGL & - & -  && 0.011 (0.001)  & 1.00 (0.00)  & 0.54 (0.04)  && 0.011 (0.001)  & 1.00 (0.00)  & 0.54 (0.04)  && 0.011 (0.001)  & 1.00 (0.00)  & 0.54 (0.04)  \\ 
  & stabJGL  &  0.166  &  0.067 && 0.015 (0.001)  & 0.88 (0.05)  & 0.66 (0.04)  && 0.015 (0.001)  & 0.90 (0.04)  & 0.66 (0.04)  && 0.015 (0.001)  & 0.91 (0.03)  & 0.66 (0.04)  \\ 
 \midrule
80 $\%$  & Glasso &  0.202  & -  && 0.026 (0.007)  & 0.40 (0.08)  & 0.50 (0.06)  && 0.018 (0.005)  & 0.57 (0.12)  & 0.48 (0.07)  && 0.015 (0.002)  & 0.75 (0.07)  & 0.55 (0.06)  \\ 
  & FGL  &  0.114  &  0.015 && 0.105 (0.030)  & 0.18 (0.06)  & 0.84 (0.04)  && 0.072 (0.023)  & 0.25 (0.08)  & 0.82 (0.04)  && 0.045 (0.012)  & 0.41 (0.10)  & 0.86 (0.03)  \\ 
  & FGL (eBIC) &  0.480  &  0.001 && 0.000 (0.000)  & -  & -  && 0.000 (0.000)  & - & -  && 0.000 (0.000)  & -  & - \\ 
  & GGL  &  0.115  &  0.008 && 0.149 (0.028)  & 0.11 (0.03)  & 0.80 (0.05)  && 0.107 (0.023)  & 0.16 (0.05)  & 0.80 (0.05)  && 0.065 (0.015)  & 0.28 (0.10)  & 0.85 (0.05)  \\ 
  & SSJGL & - & -  && 0.008 (0.001)  & 1.00 (0.01)  & 0.40 (0.04)  && 0.008 (0.001)  & 0.97 (0.03)  & 0.39 (0.04)  && 0.008 (0.001)  & 0.99 (0.01)  & 0.40 (0.04)  \\ 
  & stabJGL  &  0.166  &  0.053 && 0.014 (0.002)  & 0.84 (0.08)  & 0.59 (0.04)  && 0.012 (0.001)  & 0.90 (0.04)  & 0.54 (0.04)  && 0.012 (0.001)  & 0.93 (0.03)  & 0.56 (0.04)  \\ 
 \midrule
60 $\%$  & Glasso &  0.206  & -  && 0.026 (0.007)  & 0.41 (0.08)  & 0.51 (0.06)  && 0.017 (0.005)  & 0.59 (0.11)  & 0.48 (0.08)  && 0.015 (0.002)  & 0.75 (0.06)  & 0.55 (0.05)  \\ 
  & FGL  &  0.114  &  0.010 && 0.124 (0.029)  & 0.14 (0.04)  & 0.81 (0.03)  && 0.086 (0.022)  & 0.20 (0.05)  & 0.81 (0.04)  && 0.054 (0.014)  & 0.33 (0.08)  & 0.84 (0.04)  \\ 
  & FGL (eBIC) &  0.462  &  0.003 && 0.001 (0.003)  & 0.99 (0.05)  & 0.04 (0.12)  && 0.001 (0.002)  & 0.99 (0.02)  & 0.03 (0.10)  && 0.001 (0.002)  & 1.00 (0.01)  & 0.04 (0.11)  \\ 
  & GGL  &  0.114  &  0.006 && 0.156 (0.024)  & 0.11 (0.02)  & 0.80 (0.03)  && 0.112 (0.019)  & 0.15 (0.03)  & 0.81 (0.04)  && 0.070 (0.013)  & 0.25 (0.06)  & 0.85 (0.04)  \\ 
  & SSJGL & - & -  && 0.006 (0.001)  & 0.99 (0.02)  & 0.31 (0.03)  && 0.006 (0.001)  & 0.95 (0.03)  & 0.30 (0.03)  && 0.006 (0.001)  & 0.97 (0.03)  & 0.31 (0.03)  \\ 
  & stabJGL  &  0.166  &  0.044 && 0.015 (0.003)  & 0.75 (0.09)  & 0.55 (0.05)  && 0.012 (0.001)  & 0.87 (0.05)  & 0.50 (0.04)  && 0.011 (0.001)  & 0.92 (0.04)  & 0.52 (0.04)  \\ 
 \midrule
40 $\%$  & Glasso &  0.202  & -  && 0.027 (0.008)  & 0.39 (0.07)  & 0.51 (0.06)  && 0.018 (0.005)  & 0.57 (0.10)  & 0.49 (0.07)  && 0.015 (0.003)  & 0.77 (0.09)  & 0.55 (0.07)  \\ 
  & FGL  &  0.114  &  0.007 && 0.137 (0.024)  & 0.12 (0.03)  & 0.80 (0.04)  && 0.097 (0.021)  & 0.17 (0.04)  & 0.80 (0.04)  && 0.055 (0.013)  & 0.32 (0.07)  & 0.84 (0.05)  \\ 
  & FGL (eBIC) &  0.485  &  0.001 && 0.000 (0.000)  & - & - && 0.000 (0.000)  & -  & - && 0.000 (0.000)  & -  & - \\ 
  & GGL  &  0.114  &  0.004 && 0.158 (0.018)  & 0.10 (0.02)  & 0.81 (0.04)  && 0.115 (0.016)  & 0.14 (0.02)  & 0.81 (0.04)  && 0.067 (0.010)  & 0.26 (0.05)  & 0.85 (0.04)  \\ 
  & SSJGL & - & -  && 0.004 (0.001)  & 0.88 (0.06)  & 0.16 (0.03)  && 0.004 (0.001)  & 0.82 (0.07)  & 0.15 (0.03)  && 0.004 (0.001)  & 0.90 (0.06)  & 0.16 (0.03)  \\ 
  & stabJGL  &  0.166  &  0.038 && 0.016 (0.003)  & 0.64 (0.09)  & 0.51 (0.04)  && 0.011 (0.001)  & 0.83 (0.06)  & 0.46 (0.04)  && 0.009 (0.001)  & 0.93 (0.03)  & 0.44 (0.04)  \\ 
 \midrule
20 $\%$  & Glasso &  0.205  & -  && 0.026 (0.007)  & 0.40 (0.08)  & 0.51 (0.06)  && 0.018 (0.004)  & 0.59 (0.10)  & 0.51 (0.07)  && 0.015 (0.002)  & 0.75 (0.06)  & 0.55 (0.05)  \\ 
  & FGL  &  0.114  &  0.003 && 0.154 (0.018)  & 0.10 (0.01)  & 0.79 (0.04)  && 0.112 (0.016)  & 0.15 (0.02)  & 0.82 (0.04)  && 0.068 (0.010)  & 0.26 (0.04)  & 0.86 (0.04)  \\ 
  & FGL (eBIC) &  0.482  &  0.001 && 0.000 (0.000)  & - & -  && 0.000 (0.000)  & -   & - && 0.000 (0.00)  & -  & - \\ 
  & GGL  &  0.114  &  0.002 && 0.164 (0.011)  & 0.10 (0.01)  & 0.80 (0.04)  && 0.121 (0.010)  & 0.14 (0.01)  & 0.83 (0.03)  && 0.073 (0.007)  & 0.24 (0.02)  & 0.87 (0.03)  \\ 
  & SSJGL & - & -  && 0.003 (0.001)  & 0.83 (0.07)  & 0.12 (0.03)  && 0.003 (0.001)  & 0.81 (0.08)  & 0.12 (0.02)  && 0.003 (0.001)  & 0.90 (0.07)  & 0.13 (0.03)  \\ 
  & stabJGL  &  0.166  &  0.036 && 0.016 (0.003)  & 0.61 (0.08)  & 0.48 (0.05)  && 0.012 (0.002)  & 0.79 (0.07)  & 0.46 (0.05)  && 0.010 (0.001)  & 0.92 (0.04)  & 0.45 (0.04)  \\ 
 \midrule
0 $\%$  & Glasso &  0.204  & -  && 0.027 (0.008)  & 0.40 (0.08)  & 0.51 (0.06)  && 0.020 (0.005)  & 0.69 (0.10)  & 0.66 (0.08)  && 0.019 (0.004)  & 0.83 (0.08)  & 0.77 (0.08)  \\ 
  & FGL  &  0.114  &  0.001 && 0.165 (0.011)  & 0.10 (0.01)  & 0.81 (0.04)  && 0.118 (0.010)  & 0.16 (0.02)  & 0.94 (0.02)  && 0.069 (0.006)  & 0.28 (0.03)  & 0.97 (0.02)  \\ 
  & FGL (eBIC)  &  0.446  &  0.002 && 0.002 (0.003)  & 0.96 (0.09)  & 0.06 (0.12)  && 0.001 (0.003)  & 0.99 (0.03)  & 0.07 (0.13)  && 0.001 (0.003)  & 1.00 (0.01)  & 0.07 (0.14)  \\ 
  & GGL  &  0.114  &  0.000 && 0.168 (0.007)  & 0.10 (0.00)  & 0.81 (0.04)  && 0.121 (0.006)  & 0.16 (0.01)  & 0.94 (0.02)  && 0.071 (0.004)  & 0.28 (0.02)  & 0.97 (0.02)  \\ 
  & SSJGL & - & -  && 0.002 (0.001)  & 0.44 (0.11)  & 0.05 (0.02)  && 0.003 (0.001)  & 0.70 (0.12)  & 0.09 (0.02)  && 0.003 (0.001)  & 0.73 (0.10)  & 0.09 (0.02)  \\ 
  & stabJGL  &  0.166  &  0.024 && 0.025 (0.006)  & 0.43 (0.07)  & 0.51 (0.05)  && 0.018 (0.003)  & 0.73 (0.07)  & 0.64 (0.07)  && 0.015 (0.002)  & 0.90 (0.04)  & 0.66 (0.08)  \\ 
\bottomrule
\end{tabular}}
\end{table*}

\begin{figure}[t]
    \centering
    \includegraphics[width=0.8\textwidth]{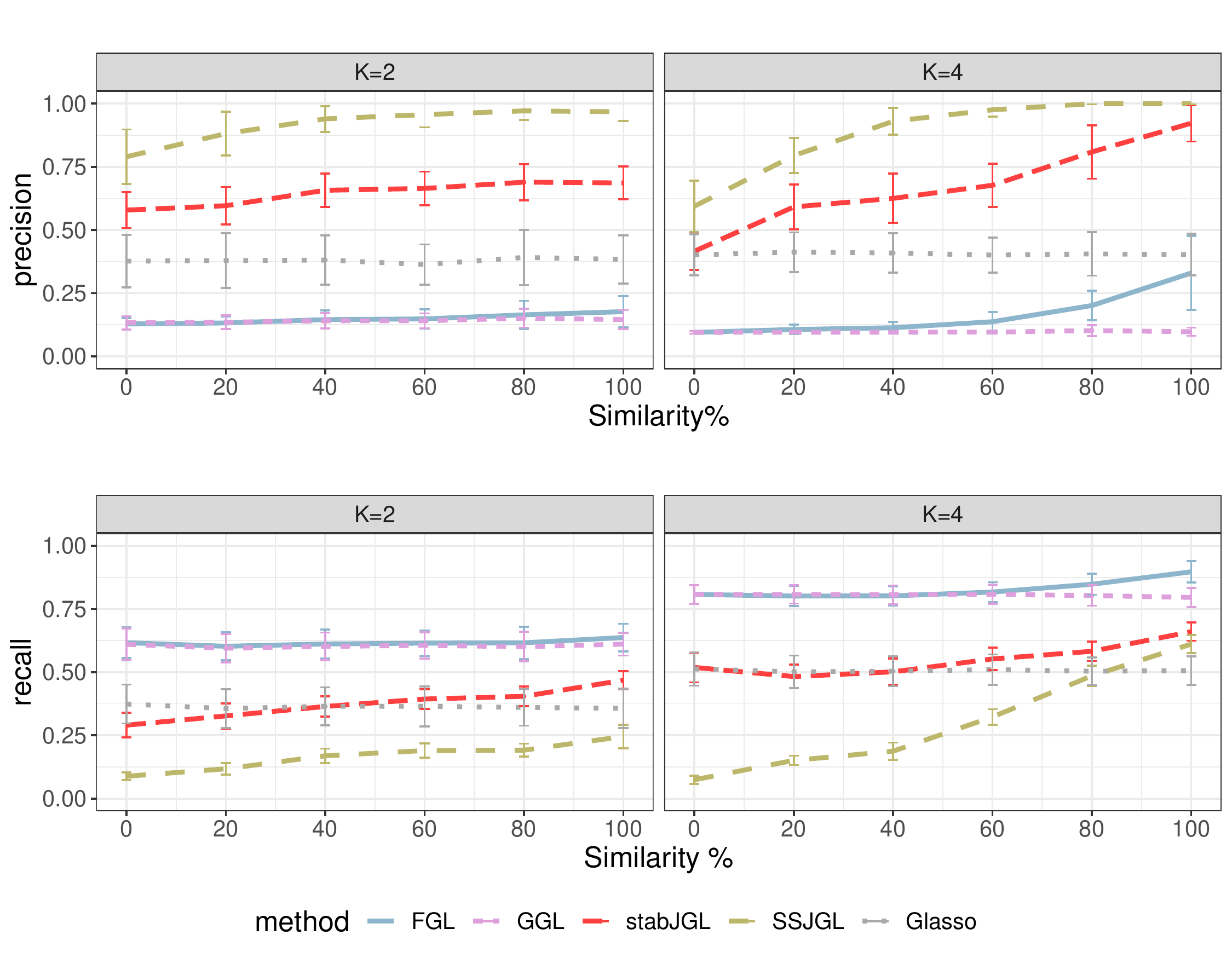}
    \caption{Performance of the Glasso, FGL and GGL tuned by AIC, SSJGL and stabJGL, reconstructing $K\in\{2, 4\}$ graphs with $p=100$ nodes and various similarity of the true graph structures. The similarity between the graphs is shown as the percentage of edges they have in common. The results are averaged over $N=100$ replicates and show the precision and recall for the first estimated graph in each setting, reconstructed from $n \in \{ 100,150\}$ observations and $n \in \{150,200,250,300\}$ observations for $K=2$ and $K=4$ respectively. Standard deviation bars are shown for all methods. All graphs have true sparsity~$0.02$. }
    \label{fig:simstudy_plot}
\end{figure}

The results for $K=2$ and $K=4$ networks are summarized in Figure \ref{fig:simstudy_plot}. The results for FGL tuned with eBIC are not shown as it did not select any edges in any of the settings. The findings from the $K=3$ case are echoed here, with FGL and GGL having high recall but very low precision and particularly GGL exhibiting a lack of adaption to increased network similarity. On the contrary, SSJGL selects very few edges and thus has high precision but very low recall, with its performance quickly deteriorating for less similar networks. StabJGL achieves a balanced precision-recall trade-off and adapts well to the level of network similarity. Consequently, stabJGL performs comparably or better than the graphical lasso depending on the degree of similarity between the networks.

A key question is whether stabJGL can achieve as high precision as the methods that give sparser networks (i.e., SSJGL) by using a lower variability threshold. Similarly, we want to see if stabJGL can achieve as high recall as the methods that infer more edges (i.e., FGL and GGL). To investigate this, we consider the same setting as in Figure \ref{fig:simstudy_plot} with $K=2$ networks, focusing specifically on the case where the two networks have $20\%$ edge agreement. Table \ref{table:simulationThresh} compares the performance of stabJGL for different values of the variability threshold $\beta_1$ to the other methods. For $\beta_1=0.01$, stabJGL gives very sparse estimates and obtains comparable precision and recall to SSJGL. For the higher threshold $\beta_1=0.2$, stabJGL selects a large number and obtains comparable recall to FGL and GGL while retaining a higher precision level. A complete comparison for all levels of edge agreement is given in the Supplement (Figure S3), where we similarly find that by varying the variability threshold $\beta_1$ we can obtain at least as high precision and/or recall as the other methods at any level of similarity. The fact that stabJGL allows the user to obtain higher or lower sparsity by changing the variability threshold means that the method can be adapted to reflect the priorities of the user (i.e., concern for false positives versus false negatives). For most applications, a middle-ground value such as $0.1$ yields a good balance between false positives and false negatives as demonstrated in the simulations.  

\begin{table}[t]
\caption{Performance of stabJGL for different values of the variability threshold $\beta_1$ on simulated data, compared to other graph reconstruction methods. The methods are used to estimate graphs with $p=100$ nodes from $K=2$ networks, both of sparsity $0.02$, of which $20\%$ of their edges are in common. The performance of stabJGL is compared to that of Glasso, FGL, GGL and SSJGL. The results are averaged over $N=100$ simulations and shows the sparsity, precision, and recall of each of the $K=2$ estimated graphs. The corresponding standard deviations are shown in parentheses. The graphs are reconstructed from $n_1=100$ and $n_2=150$ observations.
The average selected values of the penalty parameters $\lambda_1$ and $\lambda_2$ for the relevant methods is shown as well.  \label{table:simulationThresh}}
\renewcommand{\arraystretch}{1.5}
\resizebox{1\textwidth}{!}{%
\begin{tabular}{ l r r r l @{\hskip 0.03cm}l l l l @{\hskip 0.03cm}l l l}
\toprule
&&&&& \multicolumn{3}{@{}c@{}}{$n_1=100$}&&\multicolumn{3}{@{}c@{}}{$n_2=150$} \\
\cline{6-8}\cline{10-12}%
 Method & $\beta_1$ & $\lambda_1$& $\lambda_2$ &&Sparsity& Precision & Recall &&Sparsity& Precision & Recall\\
\midrule
 Glasso& - &  0.239  & -  && 0.021 (0.009)  & 0.38 (0.11)  & 0.36 (0.08)  && 0.025 (0.007)  & 0.41 (0.09)  & 0.49 (0.06)  \\ 
 FGL & - &  0.168  &  0.009 && 0.094 (0.016)  & 0.13 (0.03)  & 0.60 (0.06)  && 0.051 (0.010)  & 0.25 (0.05)  & 0.62 (0.06)  \\ 
 GGL & - &  0.167  &  0.012 && 0.091 (0.018)  & 0.14 (0.03)  & 0.59 (0.06)  && 0.049 (0.012)  & 0.26 (0.06)  & 0.61 (0.06)  \\ 
 SSJGL & -& - & -  && 0.003 (0.001)  & 0.88 (0.09)  & 0.12 (0.02)  && 0.003 (0.001)  & 0.93 (0.07)  & 0.12 (0.02)  \\ 
 stabJGL & 0.01 &  0.335  &  0.029 && 0.003 (0.002)  & 0.86 (0.13)  & 0.14 (0.07)  && 0.002 (0.001)  & 0.98 (0.04)  & 0.12 (0.06)  \\ 
  stabJGL & 0.05 &  0.271  &  0.064 && 0.006 (0.003)  & 0.81 (0.13)  & 0.24 (0.07)  && 0.004 (0.002)  & 0.96 (0.05)  & 0.20 (0.08)  \\ 
  stabJGL & 0.10 &  0.218  &  0.090 && 0.010 (0.002)  & 0.64 (0.07)  & 0.33 (0.04)  && 0.007 (0.001)  & 0.80 (0.06)  & 0.28 (0.05)  \\ 
  stabJGL & 0.20 &  0.166  &  0.093 && 0.030 (0.003)  & 0.36 (0.03)  & 0.55 (0.05)  && 0.023 (0.002)  & 0.46 (0.03)  & 0.52 (0.05)  \\ 
\bottomrule
\end{tabular}}
\end{table}

\subsection{Runtime profiling}

\begin{figure}[t]
    \centering
    \includegraphics[width=0.8\textwidth]{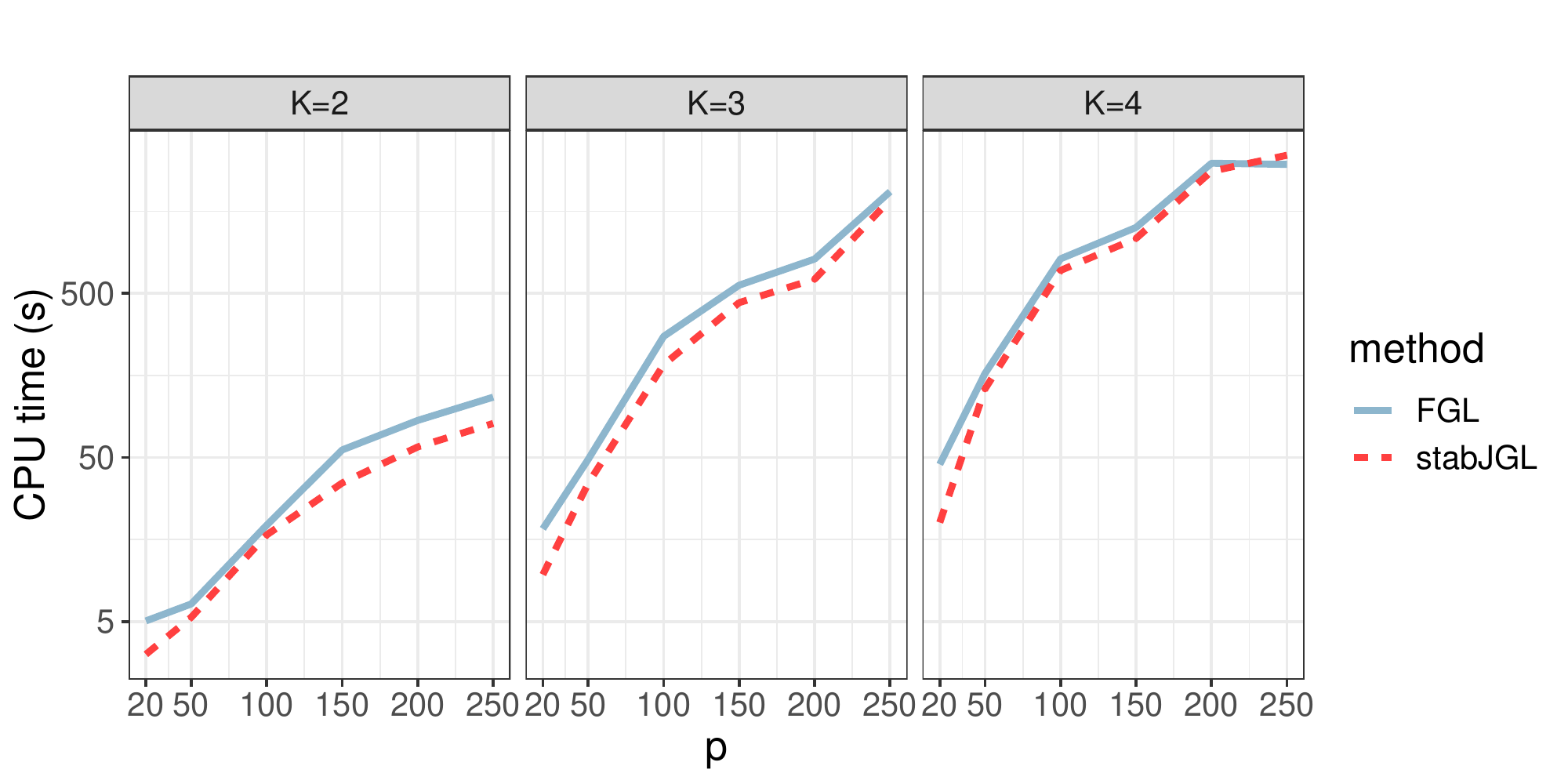}
    \caption{CPU time in seconds on a logarithmic scale used to jointly infer networks for $K \in \{2, 3, 4\}$ networks and various numbers of nodes $p$, with $n\in \{ 100,150\}$ observations, for FGL tuned with AIC and stabJGL. The computations were performed on a 16-core Intel Xeon CPU, 2.60 GHz.}
    \label{fig:time_plot}
\end{figure}

Figure \ref{fig:time_plot} shows the CPU time used to jointly infer networks for $K \in \{2, 3, 4\}$ networks and various numbers of nodes $p$, with $n\in \{ 100,150\}$ observations, for the joint graphical lasso with the fused penalty (FGL) with penalty parameters tuned with the AIC  and stabJGL with the same parameter specifications as in the previously described simulations. Due to an efficient parallelized implementation, stabJGL has an almost identical run time to FGL when the same number of $\lambda_1$ and $\lambda_2$ values are considered. Thus, the increased estimation accuracy of stabJGL does not come at a computational cost. It is important to note that due to the generalized fused lasso problem having a closed-form solution in the special case of $K=2$ \citep{danaher2014}, stabJGL is substantially faster for only two networks than for $K>2$. As stabJGL uses the fused penalty this comparison is the most relevant, but a run time comparison of all methods considered in our simulation study can be found in the Supplement (Figure S1). In the Supplement, we also demonstrate that stabJGL can be applied to problems with $p>1,000$ nodes and $K>2$ networks within reasonable time (Figure S2).  

\subsection{Pan-cancer data}

We perform a proteomic network analysis of Reverse Phase Protein Array (RPPA) data from The Cancer Genome Atlas (TCGA) across different pan-Cancer tumor types \citep{cancer2012comprehensive}. In a large proteomic pan-Cancer study of 11 TCGA tumor types, \cite{akbani2014pan} identified a major tumor super cluster consisting of hormonally responsive ``women’s cancers''. (Luminal breast cancer, ovarian cystadenocarcinoma, and uterine corpus endometrial carcinoma). Our objective is to map the proteomic network structure of the respective tumor types, so that we can get a better grasp of the common mechanisms at play in the hormonally responsive tumors. We are also interested in highlighting the differences. 

We consider mature RPPA data from Luminal breast cancer (BRCA, $n=273$), high-grade serous ovarian cystadenocarcinoma (OVCA, $n=412$), and uterine corpus endometrial carcinoma (UCEC, $n=404$). All data is downloaded from the UCSC Xena Browser \citep{UCSCXena}. The data is measured with $p=131$ high-quality antibodies that target (phospho)-proteins. To alleviate batch effects, the RPPA data is normalized with replicate-base normalization \citep{akbani2014pan}. We use stabJGL to jointly estimate the proteomic networks of the respective tumor types and interpret the results and their implications. We compare the output with that obtained with the fused joint graphical lasso (FGL) of \cite{danaher2014} with the default penalty parameter tuning with AIC as described in Subsection \ref{subsec:simdata}. Further details and code for the analysis is given at \url{https://github.com/Camiling/stabJGL_analysis}.

\subsection{Pan-cancer analysis results}

\subsubsection{Estimated proteomic networks}

\begin{figure}[t]
    \centering
    \includegraphics[width=1\textwidth]{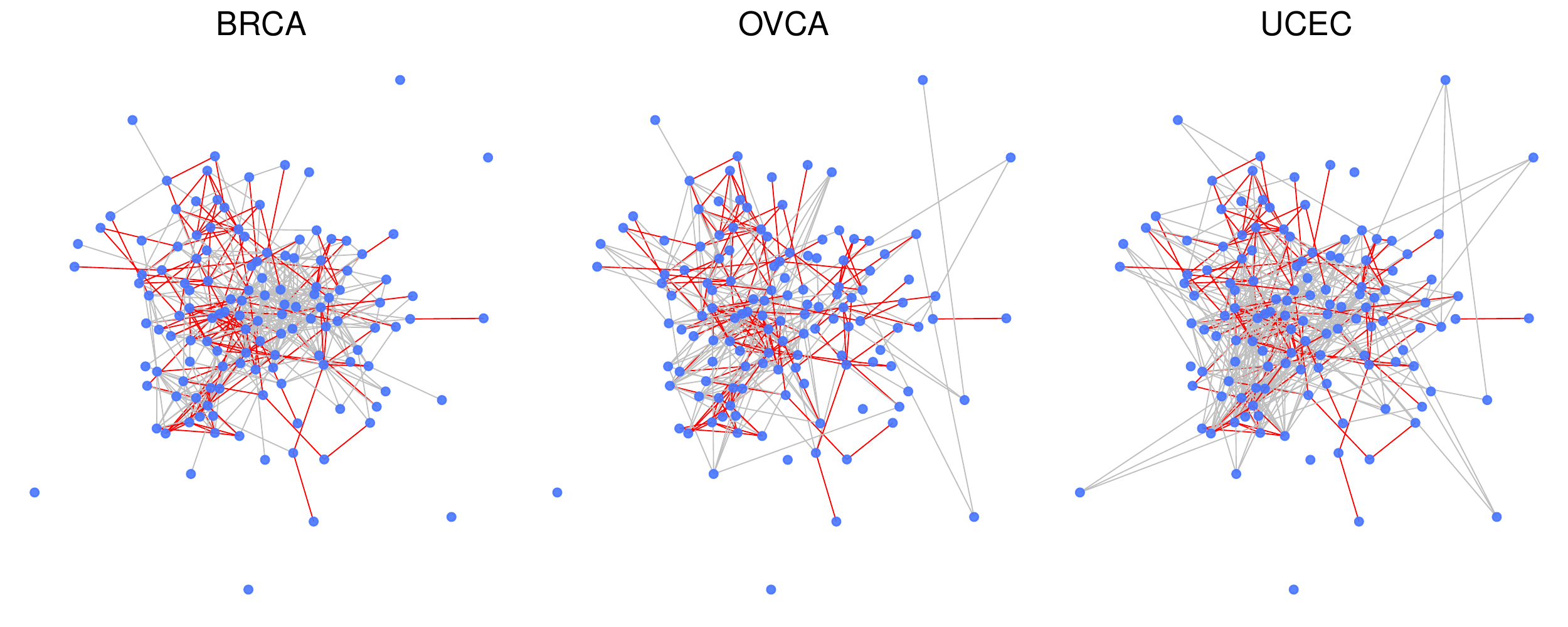}
    \caption{Proteomic network structure identified by stabJGL for the breast cancer (BRCA), ovarian cystadenocarcinoma (OVCA) and uterine corpus endometrial carcinoma (UCEC) tumors. The blue nodes represent proteins, and edges common to all three networks are marked in red, otherwise they are grey.}
    \label{fig:PanCan_all}
\end{figure}

The resulting stabJGL proteomic networks of the three tumor types are shown in Figure \ref{fig:PanCan_all}, where we observe plenty of common edges as well as network-specific ones. The sparsity as well as the selected penalty parameter values in the resulting stabJGL and FGL networks is shown in Table \ref{table:PanCanRes}. The tendency as observed in the simulations of FGL tuned by the AIC to over-select edges appears to be consistent with the findings in this context. With more than two thirds of all potential edges being determined as present by FGL, the results are challenging to interpret and derive meaningful conclusions from. From a biological standpoint, we would not expect a proteomic network to be this saturated in terms of associations due to the expected scale-free property of the degree distribution \citep{barabasi2004network}. While the degree distributions of the sparse stabJGL networks all follow a power-law with many low-degree nodes and fewer high-degree ones (hubs), an expected trait for omics data \citep{chen2004content}, the degree distributions of the FGL networks do not. 
The full degree distributions are shown in the Supplement (Figure S4).

\begin{figure}[t]
    \centering
    \includegraphics[width=0.6\textwidth]{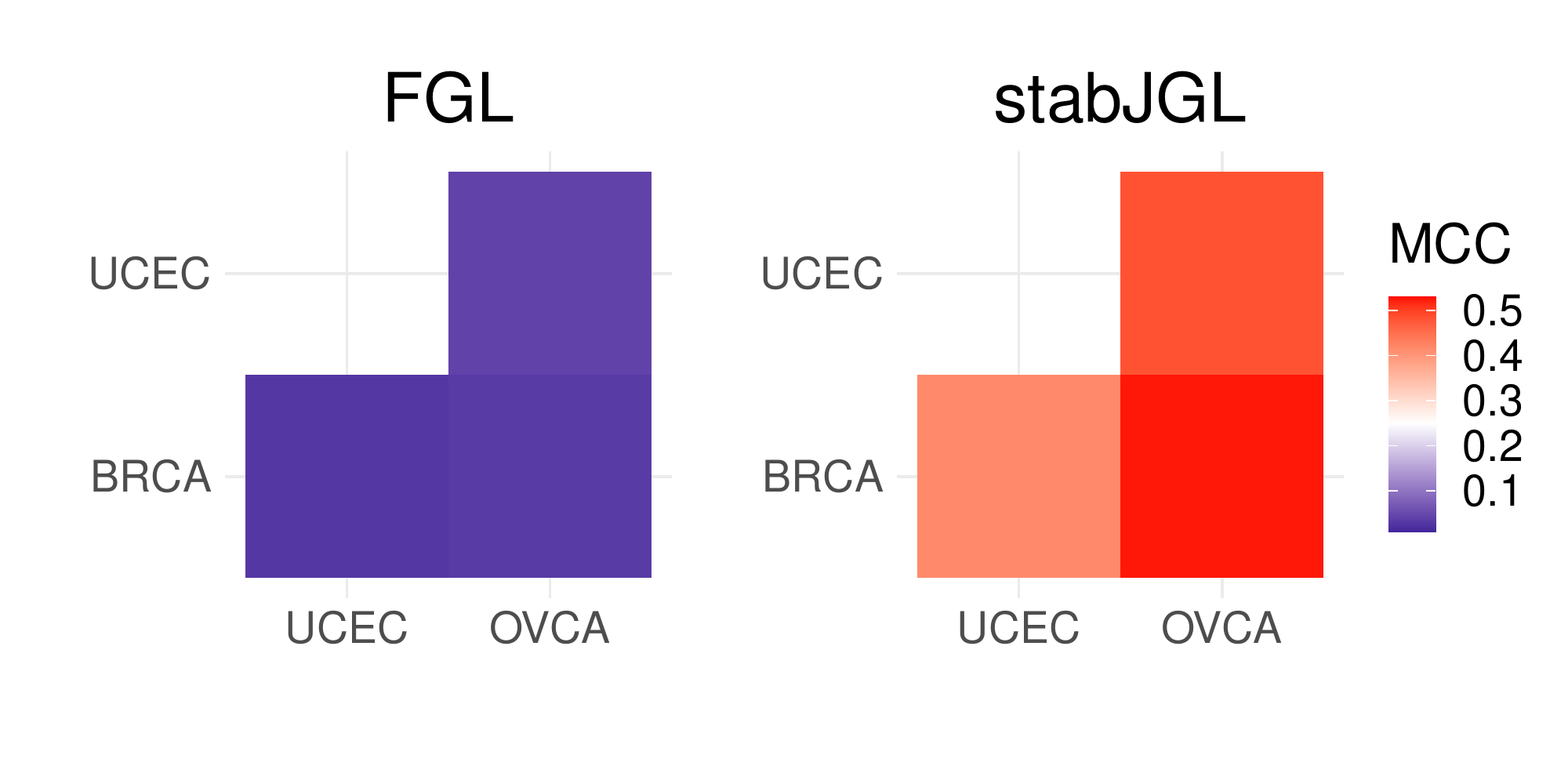}
    \caption{Pairwise Matthew's Correlation Coefficient between the proteomic network structures of the breast cancer (BRCA), ovarian cystadenocarcinoma (OVCA) and uterine corpus endometrial carcinoma (UCEC) tumors, identified by FGL tuned by the AIC and stabJGL respectively.}
    \label{fig:MCC}
\end{figure}

In terms of penalty parameters, we see that just like for the simulated data the AIC selects very small penalty parameters for FGL, resulting in little sparsity and similarity encouragement. Given the findings of \cite{akbani2014pan} about the presence of a super cluster consisting of the three hormonally responsive cancer types, it is not unreasonable to expect at least some proteomic network similarity to be encouraged by a joint method. This is achieved by stabJGL, which selects a large enough value of $\lambda_2$ to encourage similarity. A comparison of the pairwise similarities of the proteomic networks is given in Figure \ref{fig:MCC}, where similarity is measured by Matthew's Correlation Coefficient (MCC), a discretized Pearson correlation coefficient that can be used to quantify pairwise network similarities (\cite{matthews1975comparison}). StabJGL finds the networks of the three tumor types to be more similar than FGL, in accordance with the findings of \cite{akbani2014pan}.

\begin{table}
	\centering
	\caption{Network analysis results for stabJGL and FGL tuned by the AIC, applied to data from breast cancer (BRCA), ovarian cystadenocarcinoma (OVCA) and uterine corpus endometrial carcinoma (UCEC) tumors.}
	\renewcommand{\arraystretch}{1.5}
	\hspace*{-4cm}
	\begin{tabular}{l @{\hskip 0.5cm} r  r @{\hskip 0.5cm} r  r r}
		&&& \multicolumn{3}{c}{Sparsity}   \\
		\cline{4-6} 
		   & $\lambda_1$ & $\lambda_2$  & BRCA & UCEC & OVCA \\ 
		\hline
            FGL &  0.010 & 0.000 & 0.689 & 0.709  &  0.679 \\ 
            stabJGL &  0.323 & 0.008 & 0.049 &  0.036 & 0.039 \\ 
		[0.1cm] 
		\hline 
	\end{tabular}
	\label{table:PanCanRes}
	\hspace*{-4cm}
\end{table}

\subsubsection{Edge validation in STRING}

To compare the level of evidence supporting the edges detected by stabJGL and FGL tuned by the AIC in the literature, we conduct edge validation using the STRING database of known and predicted protein-protein interactions \citep{STRING}. To ensure the reliability of the validation process, we only consider the experimentally validated interactions in STRING as evidence, with default confidence score threshold $\geq 0.4$. The fraction of edges with supporting evidence in the STRING database is computed for the respective stabJGL and FGL networks and shown in Table \ref{table:string}. The analysis reveals that for all three tumor types investigated, a higher proportion of the edges detected by stabJGL had supporting evidence in the STRING database compared to those identified by FGL. 

\begin{table}
	\centering
	\caption{Comparison of evidence for edges in the respective FGL tuned by AIC and stabJGL proteomic networks of  breast cancer (BRCA), ovarian cystadenocarcinoma (OVCA) and uterine corpus endometrial carcinoma (UCEC) tumors, considering experimentally determined protein-protein interactions documented in the STRING database. The highest percentage of edges with evidence is in bold.}
	\renewcommand{\arraystretch}{1.5}
	\hspace*{-4cm}
	\begin{tabular}{l @{\hskip 0.5cm} r @{\hskip 0.5cm} r}
		& \multicolumn{2}{c}{Edge evidence $\%$} \\
		\cline{2-3}
		Data set & FGL & stabJGL\\ 
		\hline
	    BRCA  & $5.4\%$ & $\mathbf{12.3\%}$   \\ 
		UVEC & $5.6\%$ & $\mathbf{10.0\%}$\\ 
            OVCA & $5.7\%$ & $\mathbf{12.4\%}$\\ 
		[0.1cm] 
		\hline 
	\end{tabular}
	\label{table:string}
	\hspace*{-4cm}
\end{table} 

\subsubsection{Findings consistent with literature}

StabJGL successfully identifies protein-protein interactions known from literature. To highlight the findings of the proposed methodology, we only discuss edges and central proteins identified by stabJGL but not FGL. One example is the edge between activated (S345-phosphorylated) Checkpoint kinase 1 (Chk1) and DNA repair protein RAD51 homolog 1 (Rad51) in ovarian and breast cancer. The complex between the tumor suppressor BRCA2, which manifests predominantly in ovarian and breast cancer, and Rad51, is mediated by the DNA damage checkpoint Chk1 through Rad51 phosphorylation \citep{nair2020resistance, bahassi2008checkpoint}. It is also reassuring that stabJGL identifies many relevant tumor type-specific proteins as hubs in the relevant tumor type only, such as mammalian target of rapamycin (mTOR), Tuberous Sclerosis Complex 2 (Tuberin) and Ribosomal protein S6 in BRCA, all of which are involved or up/downstream of the PI3K/AKT/mTOR pathway known to frequently be deregulated in Luminal breast cancer \citep{miricescu2020pi3k}. Lists of the top hubs in the respective stabJGL and FGL networks of the different tumor types, and their node degree, are given in the Supplement (Tables S5 and S6).

\begin{figure}[t]
    \centering
    \includegraphics[width=0.75\textwidth]{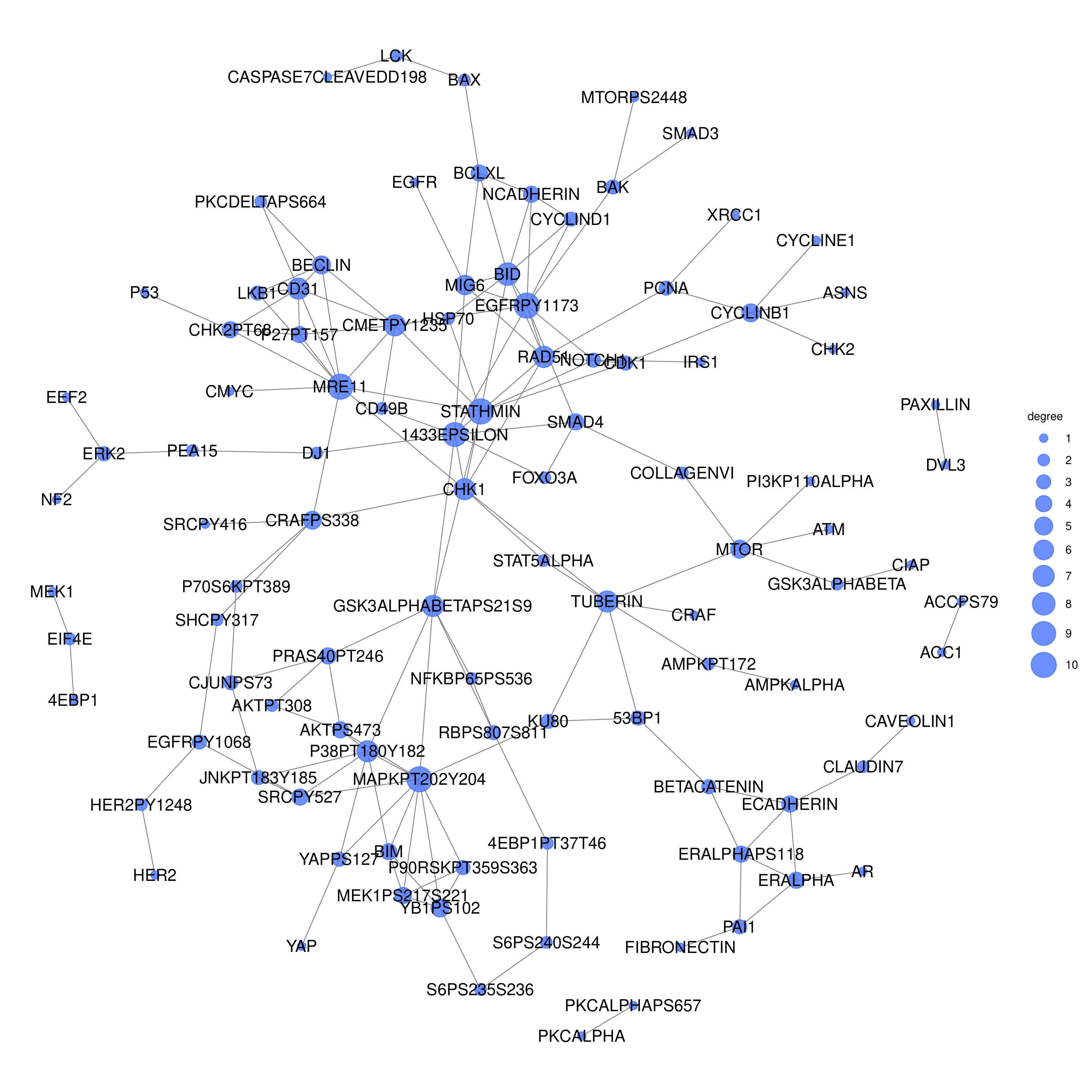}
    \caption{The proteomic network structure identified by stabJGL common to all three tumor types (BRCA, UCEC and OCVA). The node size indicates the degree in the common network structure, with proteins with more edges being represented by larger nodes.}
    \label{fig:PanCan_common}
\end{figure}

StabJGL also captures edges that we expect to be present in all three tumor types, such as the known interaction between the transcription factor Forkhead box O3 (FOXO3a) and 14-3-3-epsilon which facilitates cancer cell proliferation \citep{tzivion2011foxo, nielsen200814}. This common interaction is documented in the STRING database. Figure \ref{fig:PanCan_common} shows the network structure identified by stabJGL that is common to all three tumor types. Central proteins in this common network structure include Oncoprotein 18 (Stathmin), which is known to be relevant in all three hormonally responsive cancers due to its role in the regulation of cell growth and motility \citep{bieche1998overexpression,trovik2011stathmin, belletti2011stathmin}. 

\subsubsection{Potential candidate hubs}

The recovery of documented links in the protein networks estimated by stabJGL highlights its capability to detect numerous relevant proteins and interactions. The potential for new discoveries is however an important aspect of stabJGL, as suggested by its good performance on simulated data. For example, stabJGL identifies phosphorylated epidermal growth factor receptor (EGFR) as a central hub protein in all three tumor types. While known to be relevant in ovarian cancer \cite{zhang2016integrated,yang2013predicting}, the role of activated EGFR in uterine corpus endometrial carcinoma and Luminal breast cancer and is not yet clarified. Our findings suggest it could be relevant in all three hormonally responsive tumor types. Further, Platelet endothelial cell adhesion molecule (CD31) is found to be a central protein in UCEC only. The protein is important for angiogenesis and has been implicated in other tumor types such as haemangioma \citep{bergom2005mechanisms}. Its prominence in the proteomic UCEC network suggests it may play a crucial role in this tumor type as well. Overall, these results showcase how stabJGL can aid in generating hypotheses by identifying central proteins and associations.


\section{Discussion}
\label{sec:discussion}

Suitable sparsity and similarity selection is key for capturing and studying multiple related biological networks. We have proposed the stabJGL algorithm, which determines the penalty parameters in the fused graphical lasso for multiple networks based on the principle of model stability. StabJGL demonstrably avoids the under- or over-selection of edges observed in state-of-the-art selection methods based on information criteria, 
and succeeds at leveraging network similarities to a suitable degree. 
Consequently, the method can be employed in situations where the actual degree of similarity is uncertain, resulting in marked benefits with minimal risks associated with its use. StabJGL offers a fast parallelized implementation, particularly for  $K=2$ networks as a closed-form solution exists. We successfully apply the method to problems with $p>1,000$ nodes and $K>2$ networks. 

With our novel approach, we can identify both common and distinct mechanisms in the proteomic networks of different types of hormonally responsive women's cancers. The results obtained with stabJGL are in line with known biology and compliment those of \cite{akbani2014pan} by offering additional understanding of the underlying mechanisms in action. By recognizing various proteins as highly critical in the proteomic networks, stabJGL suggests their possible involvement in driving the diseases. The method both identifies proteins that are central in all three hormonally responsive cancers (e.g., phosphorylated EGFR) and proteins of tumor-specific relevance (e.g., CD31 in UCEC). 

Future extensions of the method can include alternative measures of variability, such as the entropy (see, e.g., \cite{lartigue2020gaussian}). Further, while the method is formulated specifically for the joint graphical lasso with the fused penalty, it can in principle be used for any joint network approach requiring the tuning of sparsity- and similarity-controlling parameters. One potential method of application is JCGL \citep{huang2017joint}, which is based on a group lasso penalty and currently fixes the penalty parameters according to theoretical results. 

To conclude, stabJGL provides a reliable approach to joint network inference of omics data. The output can provide a better understanding of both common and data type-specific mechanisms, which can be used for hypothesis generation regarding potential therapeutic targets.

\section*{Software}
 A user-friendly R package for stabJGL with tutorials is available on Github: \url{https://github.com/Camiling/stabJGL}. R code for the simulations and data analyses in this paper is available at \url{https://github.com/Camiling/stabJGL_simulations} and \url{https://github.com/Camiling/stabJGL_analysis}.


\section*{Funding}

This research is funded by the UK Medical Research Council programme MRC MC UU 00002/10 (C.L. and S.R.) and Aker Scholarship (C.L.).



\bibliographystyle{unsrtnat} 
\bibliography{bibliography.bib}       

\appendix
\renewcommand\thefigure{\thesection.\arabic{figure}}    
\renewcommand\thetable{\thesection.\arabic{table} }
\section{Appendix}
\setcounter{figure}{0}
\setcounter{table}{0}

\section{Runtime profiling}
\label{Supp:sec:runtime}

\begin{figure*}[t]
    \centering
    \includegraphics[width=\textwidth]{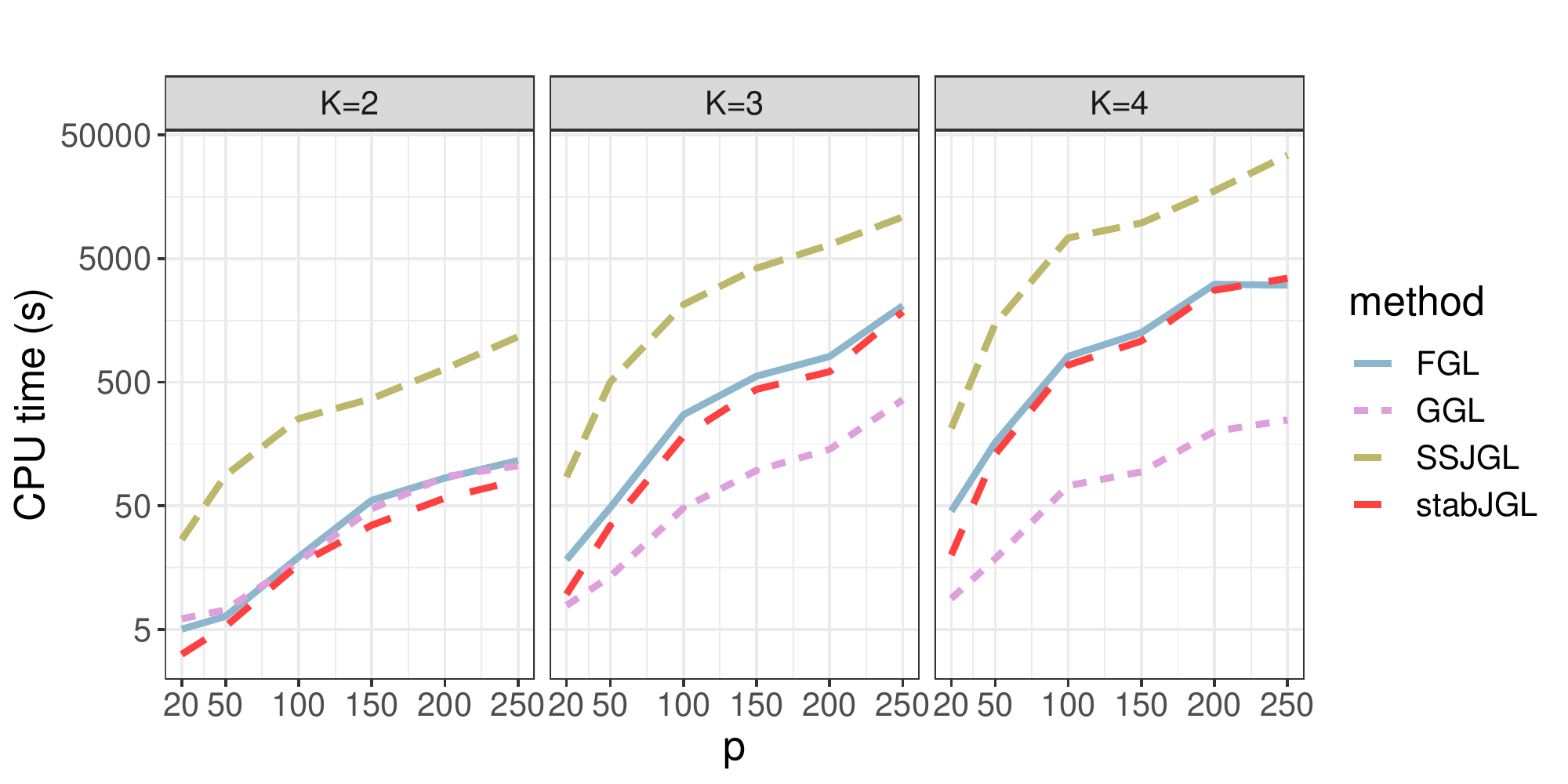}
    \caption{CPU time in seconds on a logarithmic scale used to jointly infer networks for $K \in \{2, 4\}$ networks and various numbers of nodes $p$, with $n\in \{ 100,150\}$ observations, for the fused and group joint graphical lasso with AIC penalty parameter selection (FGL and GGL), the Bayesian spike-and-slab joint graphical lasso (SSJGL) and stabJGL. The computations were performed on a 16-core Intel Xeon CPU, 2.60 GHz.}
    \label{fig:time_plot_extended}
\end{figure*}

\begin{figure*}[t]
    \centering
    \includegraphics[width=\textwidth]{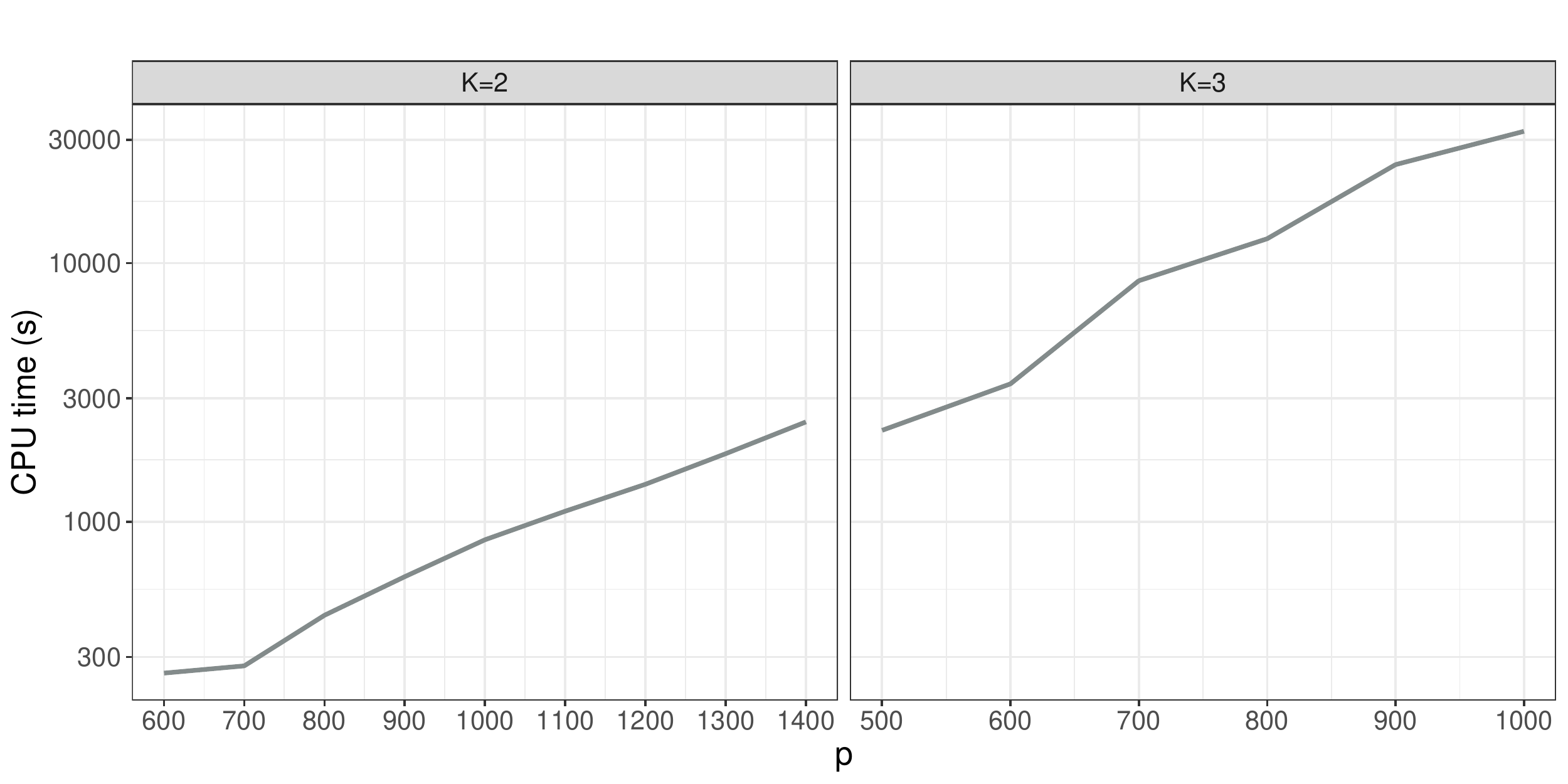}
    \caption{CPU time in seconds on a logarithmic scale used by stabJGL to jointly infer (a) $K=2$ networks with various numbers of nodes $p$ from $n_1=n_2=500$ observations and (b) $K=3$ networks with various numbers of nodes $p$ from $n_1=n_2=n_3=500$ observations. The computations were performed on a 16-core Intel Xeon CPU, 2.60 GHz.}
    \label{fig:time_plot_extended_stabJGL}
\end{figure*}

Figure \ref{fig:time_plot_extended} compares the time used to infer $K \in \{2, 4\}$ networks with various numbers of nodes $p$, for the different network reconstruction methods. We only consider the AIC selection for FGL as the eBIC considers the same grid of values and hence has identical running time. All methods are run with the same parameter specifications as in the main simulation study. The simulated networks are set to have $50\%$ of their edges in common, generated with the same approach as in the main simulation study. As discussed by \cite{danaher2014}, the group joint graphical lasso is faster that its fused counterpart. The Bayesian spike-and-slab joint graphical lasso is substantially slower than the other methods, taking around ten times longer than the fused joint graphical lasso and stabJGL. 
Figure \ref{fig:time_plot_extended_stabJGL} shows the time used by stabJGL to infer $K \in \{2, 3\}$ networks with various numbers of nodes $p$ and $50\%$ of their edges in common. We see that for $K=2$ networks, inference for $p=1,400$ nodes is feasible within half an hour, while for $K=3$ inference for $p=1,000$ nodes is feasible within about eight hours. As discussed by \cite{danaher2014}, there is an explicit solution to the fused joint graphical lasso problem for $K=2$ and hence inference is much faster for stabJGL as well in that case. 

\section{Additional simulation scenarios}
\label{Supp:sec:sim}

Additional simulation studies are conducted to assess a wider range of scenarios and compare the performance of the graphical lasso (Glasso), the fused joint graphical lasso (FGL) and the group joint graphical lasso (GGL) tuned by AIC, the fused joint graphical lasso tuned by eBIC, the Bayesian spike-and-slab joint graphical lasso (SSJGL) and stabJGL. Table \ref{table:simulationA} shows the network reconstruction performance of the methods in a $K=3$ network setting with $p=200$ nodes and various similarity of the true graph structures, averaged over $N=100$ simulations. Similarly, Table \ref{table:simulationB} shows the results for a $K=4$ setting with $p=100$ nodes. In Table \ref{table:simulationC}, the results are shown for a $K=2$ network setting with $p=100$ nodes. Finally, the results from a $K=2$ network setting with $p=300$ nodes are shown in Table \ref{table:simulationD}. In the latter case, due to the longer run time of SSJGL as demonstrated in Section \ref{Supp:sec:runtime}, this method is omitted to make the simulation study feasible within reasonable time ($<48$ hours).
The results from the additional simulation are in line with those from the main simulation study; stabJGL succeeds at capturing both the sparsity level and similarity between the networks to a better degree than FGL and GGl, while either outperforming the standard graphical lasso for highly similar networks or getting comparable results for unrelated networks. FGL with the alternative eBIC selection mostly selects empty graphs. Finally, SSJGL select very few edges, leading to high precision but very low recall in all cases.

\begin{sidewaystable}
\caption{Performance of the different graph reconstruction methods in simulations, reconstructing graphs with $p=200$ nodes from $K=3$ classes with various similarity of the true graph structures. The methods included are the graphical lasso (Glasso), the fused joint graphical lasso tuned by the AIC (FGL) and by the extended BIC (eBIC), the group joint graphical lasso (GGL), the Bayesian spike-and-slab joint graphical lasso (SSJGL) and stabJGL. The similarity (percentage of edges that are in common) of the graphs is shown. The results are averaged over $N=100$ simulations and shows the sparsity, precision, and recall of each of the $K=3$ estimated graphs. The corresponding standard deviations are shown as well. The graphs are reconstructed from $n_1=150$, $n_2=200$ and $n_3=300$ observations. All graphs have sparsity $0.01$. 
The average selected values of the penalty parameters $\lambda_1$ and $\lambda_2$ for the relevant methods is shown as well.  \label{table:simulationA}}
\renewcommand{\arraystretch}{1.2}
\hspace*{-2cm}
\begin{tabular}{r l r r l @{\hskip 0.1cm}l l l l @{\hskip 0.1cm}l l l l@{\hskip 0.1cm}l l l}
\toprule
&&&&& \multicolumn{3}{@{}c@{}}{$n_1=150$}&&\multicolumn{3}{@{}c@{}}{$n_2=200$}&&\multicolumn{3}{@{}c@{}}{$n_3=300$} \\
\cline{6-8}\cline{10-12} \cline{14-16}%
Similarity & Method & $\lambda_1$& $\lambda_2$ &&Sparsity& Precision & Recall &&Sparsity& Precision & Recall&&Sparsity& Precision & Recall\\
\midrule
100 $\%$ & Glasso &  0.201  & -  && 0.018 (0.007)  & 0.22 (0.05)  & 0.37 (0.05)  && 0.010 (0.004)  & 0.40 (0.09)  & 0.37 (0.05)  && 0.006 (0.002)  & 0.64 (0.09)  & 0.39 (0.05)  \\ 
  & FGL  &  0.166  &  0.013 && 0.027 (0.012)  & 0.21 (0.11)  & 0.49 (0.04)  && 0.015 (0.006)  & 0.38 (0.15)  & 0.48 (0.03)  && 0.007 (0.002)  & 0.68 (0.14)  & 0.46 (0.04)  \\ 
  & FGL (eBIC) &  0.488  &  0.000 && 0.000 (0.000)  & -  & -  && 0.000 (0.000)  & -  & - && 0.000 (0.000)  & -  & - \\ 
  & GGL  &  0.166  &  0.004 && 0.044 (0.006)  & 0.12 (0.02)  & 0.50 (0.04)  && 0.024 (0.004)  & 0.21 (0.04)  & 0.49 (0.03)  && 0.010 (0.002)  & 0.48 (0.06)  & 0.48 (0.04)  \\ 
  & SSJGL & - & -  && 0.002 (0.000)  & 1.00 (0.00)  & 0.25 (0.03)  && 0.002 (0.000)  & 1.00 (0.00)  & 0.25 (0.03)  && 0.002 (0.000)  & 1.00 (0.00)  & 0.25 (0.03)  \\ 
  & stabJGL  &  0.166  &  0.078 && 0.005 (0.000)  & 0.89 (0.05)  & 0.44 (0.02)  && 0.005 (0.000)  & 0.90 (0.03)  & 0.44 (0.02)  && 0.005 (0.000)  & 0.91 (0.03)  & 0.44 (0.02)  \\ 
 \midrule
80 $\%$  & Glasso &  0.200  & -  && 0.017 (0.006)  & 0.23 (0.06)  & 0.36 (0.05)  && 0.010 (0.003)  & 0.40 (0.09)  & 0.38 (0.05)  && 0.006 (0.002)  & 0.63 (0.10)  & 0.38 (0.05)  \\ 
  & FGL  &  0.166  &  0.005 && 0.039 (0.011)  & 0.14 (0.05)  & 0.48 (0.04)  && 0.021 (0.006)  & 0.25 (0.08)  & 0.49 (0.04)  && 0.008 (0.002)  & 0.56 (0.13)  & 0.45 (0.04)  \\ 
  & FGL (eBIC) &  0.489  &  0.000 && 0.000 (0.000)  & -  & -  && 0.000 (0.000)  & -  & - && 0.000 (0.000)  & -  & - \\ 
  & GGL  &  0.166  &  0.003 && 0.045 (0.005)  & 0.11 (0.01)  & 0.49 (0.03)  && 0.025 (0.003)  & 0.21 (0.03)  & 0.50 (0.04)  && 0.010 (0.001)  & 0.49 (0.06)  & 0.46 (0.03)  \\ 
  & SSJGL & - & -  && 0.002 (0.000)  & 1.00 (0.00)  & 0.17 (0.02)  && 0.002 (0.000)  & 0.98 (0.02)  & 0.17 (0.02)  && 0.002 (0.000)  & 0.99 (0.01)  & 0.17 (0.02)  \\ 
  & stabJGL  &  0.166  &  0.064 && 0.005 (0.001)  & 0.84 (0.07)  & 0.37 (0.03)  && 0.004 (0.000)  & 0.88 (0.04)  & 0.35 (0.03)  && 0.004 (0.000)  & 0.91 (0.04)  & 0.34 (0.03)  \\ 
 \midrule
60 $\%$  & Glasso &  0.196  & -  && 0.018 (0.006)  & 0.22 (0.06)  & 0.37 (0.05)  && 0.011 (0.004)  & 0.37 (0.10)  & 0.36 (0.05)  && 0.007 (0.002)  & 0.60 (0.10)  & 0.41 (0.06)  \\ 
  & FGL  &  0.166  &  0.003 && 0.043 (0.008)  & 0.12 (0.03)  & 0.50 (0.04)  && 0.022 (0.005)  & 0.22 (0.05)  & 0.47 (0.04)  && 0.009 (0.002)  & 0.52 (0.08)  & 0.46 (0.04)  \\ 
  & FGL (eBIC) &  0.484  &  0.000 && 0.000 (0.000)  & -  & -  && 0.000 (0.000)  & -  & - && 0.000 (0.000)  & -  & - \\ 
  & GGL  &  0.166  &  0.003 && 0.045 (0.005)  & 0.11 (0.01)  & 0.50 (0.04)  && 0.024 (0.003)  & 0.20 (0.02)  & 0.47 (0.03)  && 0.009 (0.001)  & 0.50 (0.05)  & 0.46 (0.04)  \\ 
  & SSJGL & - & -  && 0.001 (0.000)  & 1.00 (0.01)  & 0.10 (0.02)  && 0.001 (0.000)  & 0.95 (0.05)  & 0.10 (0.02)  && 0.001 (0.000)  & 0.99 (0.02)  & 0.10 (0.02)  \\ 
  & stabJGL  &  0.166  &  0.056 && 0.004 (0.001)  & 0.77 (0.07)  & 0.34 (0.03)  && 0.003 (0.000)  & 0.85 (0.04)  & 0.29 (0.03)  && 0.003 (0.000)  & 0.92 (0.03)  & 0.29 (0.03)  \\ 
 \midrule
40 $\%$  & Glasso &  0.199  & -  && 0.017 (0.006)  & 0.23 (0.05)  & 0.36 (0.05)  && 0.011 (0.004)  & 0.38 (0.11)  & 0.36 (0.06)  && 0.006 (0.002)  & 0.63 (0.13)  & 0.37 (0.07)  \\ 
  & FGL  &  0.166  &  0.002 && 0.046 (0.008)  & 0.11 (0.02)  & 0.50 (0.04)  && 0.024 (0.005)  & 0.21 (0.04)  & 0.48 (0.04)  && 0.009 (0.002)  & 0.50 (0.07)  & 0.44 (0.04)  \\ 
  & FGL (eBIC) &  0.482  &  0.000 && 0.000 (0.000)  & -  & -  && 0.000 (0.000)  & -  & - && 0.000 (0.000)  & -  & - \\ 
  & GGL  &  0.166  &  0.002 && 0.046 (0.005)  & 0.11 (0.01)  & 0.50 (0.03)  && 0.024 (0.003)  & 0.20 (0.02)  & 0.48 (0.04)  && 0.009 (0.001)  & 0.49 (0.05)  & 0.44 (0.03)  \\ 
  & SSJGL & - & -  && 0.001 (0.000)  & 0.98 (0.04)  & 0.06 (0.01)  && 0.001 (0.000)  & 0.92 (0.07)  & 0.05 (0.01)  && 0.001 (0.000)  & 0.98 (0.04)  & 0.06 (0.01)  \\ 
  & stabJGL  &  0.166  &  0.057 && 0.004 (0.001)  & 0.75 (0.09)  & 0.28 (0.03)  && 0.003 (0.000)  & 0.83 (0.05)  & 0.23 (0.03)  && 0.002 (0.000)  & 0.91 (0.04)  & 0.23 (0.03)  \\ 
 \midrule
20 $\%$  & Glasso &  0.198  & -  && 0.019 (0.007)  & 0.22 (0.06)  & 0.38 (0.05)  && 0.010 (0.004)  & 0.38 (0.11)  & 0.34 (0.07)  && 0.006 (0.002)  & 0.63 (0.11)  & 0.39 (0.06)  \\ 
  & FGL  &  0.166  &  0.001 && 0.047 (0.005)  & 0.11 (0.01)  & 0.50 (0.03)  && 0.024 (0.003)  & 0.20 (0.03)  & 0.47 (0.04)  && 0.010 (0.001)  & 0.48 (0.06)  & 0.47 (0.03)  \\ 
  & FGL (eBIC) &  0.486  &  0.000 && 0.000 (0.000)  & -  & -  && 0.000 (0.000)  & -  & - && 0.000 (0.000)  & -  & - \\ 
  & GGL  &  0.166  &  0.002 && 0.046 (0.005)  & 0.11 (0.01)  & 0.50 (0.03)  && 0.024 (0.003)  & 0.20 (0.03)  & 0.47 (0.04)  && 0.010 (0.001)  & 0.48 (0.05)  & 0.47 (0.03)  \\ 
  & SSJGL & - & -  && 0.000 (0.000)  & 0.90 (0.12)  & 0.03 (0.01)  && 0.000 (0.000)  & 0.87 (0.12)  & 0.03 (0.01)  && 0.000 (0.000)  & 0.96 (0.07)  & 0.03 (0.01)  \\ 
  & stabJGL  &  0.166  &  0.052 && 0.004 (0.001)  & 0.65 (0.09)  & 0.27 (0.03)  && 0.003 (0.000)  & 0.82 (0.05)  & 0.22 (0.03)  && 0.002 (0.000)  & 0.91 (0.04)  & 0.22 (0.03)  \\ 
 \midrule
0 $\%$  & Glasso &  0.201  & -  && 0.017 (0.005)  & 0.23 (0.05)  & 0.36 (0.04)  && 0.011 (0.004)  & 0.54 (0.12)  & 0.57 (0.08)  && 0.008 (0.003)  & 0.73 (0.12)  & 0.52 (0.09)  \\ 
  & FGL  &  0.140  &  0.005 && 0.077 (0.031)  & 0.08 (0.02)  & 0.56 (0.07)  && 0.048 (0.023)  & 0.20 (0.08)  & 0.79 (0.06)  && 0.023 (0.013)  & 0.40 (0.16)  & 0.73 (0.09)  \\ 
  & FGL (eBIC) &  0.497  &  0.001 && 0.000 (0.000)  & -  & -  && 0.000 (0.000)  & -  & - && 0.000 (0.000)  & -  & - \\ 
  & GGL  &  0.166  &  0.000 && 0.048 (0.002)  & 0.10 (0.01)  & 0.50 (0.03)  && 0.027 (0.001)  & 0.28 (0.01)  & 0.74 (0.03)  && 0.012 (0.001)  & 0.56 (0.03)  & 0.65 (0.03)  \\ 
  & SSJGL & - & -  && 0.000 (0.000)  & 0.47 (0.21)  & 0.01 (0.01)  && 0.000 (0.000)  & 0.79 (0.18)  & 0.02 (0.01)  && 0.000 (0.000)  & 0.84 (0.17)  & 0.02 (0.01)  \\ 
  & stabJGL  &  0.166  &  0.047 && 0.005 (0.001)  & 0.53 (0.09)  & 0.23 (0.04)  && 0.004 (0.001)  & 0.84 (0.05)  & 0.31 (0.06)  && 0.003 (0.000)  & 0.92 (0.04)  & 0.24 (0.04)  \\ 
\bottomrule
\end{tabular}
\end{sidewaystable}

\begin{sidewaystable}
\caption{Performance of the different graph reconstruction methods in simulations, reconstructing graphs with $p=100$ nodes from $K=4$ classes with various similarity of the true graph structures. The methods included are the graphical lasso (Glasso), the fused joint graphical lasso tuned by the AIC (FGL) and by the extended BIC (eBIC), the group joint graphical lasso (GGL), the Bayesian spike-and-slab joint graphical lasso (SSJGL) and stabJGL. The similarity (percentage of edges that are in common) of the graphs is shown. The results are averaged over $N=100$ simulations and shows the sparsity, precision, and recall of each of the $K=4$ estimated graphs. The corresponding standard deviations are shown as well. The graphs are reconstructed from $n_1=150$, $n_2=200$, $n_3=250$ and $n_4=300$ observations. All graphs have sparsity $0.02$. 
The average selected values of the penalty parameters $\lambda_1$ and $\lambda_2$ for the relevant methods is shown as well.  \label{table:simulationB}}
\renewcommand{\arraystretch}{1.3}
\hspace*{-2cm}
\resizebox{1.1\textwidth}{!}{%
\begin{tabular}{r l r r l @{\hskip 0.1cm}l l l l @{\hskip 0.1cm}l l l l @{\hskip 0.1cm}l l l l@{\hskip 0.1cm}l l l}
\toprule
&&&&& \multicolumn{3}{@{}c@{}}{$n_1=150$}&&\multicolumn{3}{@{}c@{}}{$n_2=200$}&&\multicolumn{3}{@{}c@{}}{$n_3=250$}&&\multicolumn{3}{@{}c@{}}{$n_4=300$} \\
\cline{6-8}\cline{10-12} \cline{14-16} \cline{18-20}%
Similarity & Method & $\lambda_1$& $\lambda_2$ &&Sparsity& Precision & Recall && Sparsity& Precision & Recall &&Sparsity& Precision & Recall&&Sparsity& Precision & Recall\\
\midrule
100 $\%$ & Glasso &  0.206  & -  && 0.026 (0.007)  & 0.40 (0.08)  & 0.51 (0.06)  && 0.018 (0.005)  & 0.57 (0.09)  & 0.51 (0.06)  && 0.016 (0.003)  & 0.69 (0.08)  & 0.53 (0.06)  && 0.015 (0.003)  & 0.75 (0.08)  & 0.55 (0.06)  \\ 
  & FGL  &  0.114  &  0.02 && 0.065 (0.028)  & 0.33 (0.15)  & 0.90 (0.04)  && 0.047 (0.020)  & 0.44 (0.15)  & 0.92 (0.03)  && 0.038 (0.014)  & 0.53 (0.14)  & 0.92 (0.04)  && 0.033 (0.010)  & 0.61 (0.13)  & 0.93 (0.03)  \\ 
  & FGL (eBIC) &  0.232  &  0.036 && 0.008 (0.002)  & 0.98 (0.04)  & 0.38 (0.09)  && 0.008 (0.002)  & 0.99 (0.02)  & 0.38 (0.09)  && 0.008 (0.002)  & 1.00 (0.01)  & 0.38 (0.09)  && 0.007 (0.002)  & 1.00 (0.00)  & 0.37 (0.09)  \\ 
  & GGL  &  0.114  &  0.002 && 0.165 (0.014)  & 0.10 (0.02)  & 0.80 (0.04)  && 0.123 (0.011)  & 0.14 (0.02)  & 0.84 (0.04)  && 0.095 (0.010)  & 0.18 (0.04)  & 0.86 (0.04)  && 0.075 (0.007)  & 0.24 (0.03)  & 0.88 (0.04)  \\ 
  & SSJGL & - & -  && 0.012 (0.001)  & 1.00 (0.00)  & 0.61 (0.04)  && 0.012 (0.001)  & 1.00 (0.00)  & 0.61 (0.04)  && 0.012 (0.001)  & 1.00 (0.00)  & 0.61 (0.04)  && 0.012 (0.001)  & 1.00 (0.00)  & 0.61 (0.04)  \\ 
  & stabJGL  &  0.166  &  0.042 && 0.014 (0.002)  & 0.92 (0.07)  & 0.66 (0.04)  && 0.014 (0.001)  & 0.95 (0.04)  & 0.66 (0.04)  && 0.014 (0.001)  & 0.97 (0.02)  & 0.66 (0.04)  && 0.014 (0.001)  & 0.97 (0.02)  & 0.66 (0.04)  \\ 
 \midrule
80 $\%$ & Glasso &  0.201  & -  && 0.026 (0.007)  & 0.41 (0.09)  & 0.50 (0.05)  && 0.018 (0.005)  & 0.57 (0.12)  & 0.49 (0.07)  && 0.016 (0.003)  & 0.70 (0.09)  & 0.54 (0.07)  && 0.015 (0.003)  & 0.75 (0.07)  & 0.55 (0.06)  \\ 
  & FGL  &  0.114  &  0.012 && 0.091 (0.025)  & 0.20 (0.06)  & 0.85 (0.04)  && 0.061 (0.019)  & 0.30 (0.09)  & 0.83 (0.04)  && 0.047 (0.015)  & 0.40 (0.11)  & 0.86 (0.03)  && 0.038 (0.010)  & 0.48 (0.11)  & 0.87 (0.03)  \\ 
  & FGL (eBIC)  &  0.407  &  0.008 && 0.003 (0.004)  & 0.97 (0.06)  & 0.13 (0.18)  && 0.002 (0.003)  & 0.99 (0.02)  & 0.11 (0.16)  && 0.002 (0.003)  & 1.00 (0.01)  & 0.12 (0.17)  && 0.002 (0.003)  & 1.00 (0.00)  & 0.12 (0.17)  \\ 
  & GGL  &  0.114  &  0.003 && 0.161 (0.019)  & 0.10 (0.02)  & 0.80 (0.04)  && 0.117 (0.016)  & 0.14 (0.04)  & 0.81 (0.04)  && 0.090 (0.012)  & 0.19 (0.04)  & 0.84 (0.03)  && 0.072 (0.010)  & 0.24 (0.05)  & 0.86 (0.04)  \\ 
  & SSJGL & - & -  && 0.010 (0.001)  & 1.00 (0.00)  & 0.49 (0.04)  && 0.010 (0.001)  & 0.94 (0.02)  & 0.45 (0.03)  && 0.010 (0.001)  & 0.96 (0.02)  & 0.46 (0.04)  && 0.010 (0.001)  & 0.96 (0.02)  & 0.47 (0.04)  \\ 
  & stabJGL  &  0.166  &  0.031 && 0.015 (0.002)  & 0.81 (0.11)  & 0.58 (0.04)  && 0.012 (0.001)  & 0.92 (0.05)  & 0.54 (0.04)  && 0.012 (0.001)  & 0.96 (0.03)  & 0.56 (0.04)  && 0.011 (0.001)  & 0.97 (0.02)  & 0.55 (0.04)  \\ 
 \midrule
60 $\%$ & Glasso &  0.203  & -  && 0.026 (0.007)  & 0.40 (0.07)  & 0.51 (0.06)  && 0.018 (0.005)  & 0.57 (0.10)  & 0.49 (0.07)  && 0.015 (0.003)  & 0.70 (0.08)  & 0.53 (0.05)  && 0.014 (0.003)  & 0.75 (0.08)  & 0.53 (0.06)  \\ 
  & FGL  &  0.114  &  0.006 && 0.127 (0.030)  & 0.14 (0.04)  & 0.82 (0.04)  && 0.087 (0.024)  & 0.20 (0.06)  & 0.80 (0.04)  && 0.068 (0.019)  & 0.27 (0.08)  & 0.83 (0.04)  && 0.052 (0.015)  & 0.35 (0.10)  & 0.83 (0.04)  \\ 
  & FGL (eBIC) &  0.435  &  0.005 && 0.002 (0.004)  & 0.97 (0.07)  & 0.08 (0.16)  && 0.002 (0.003)  & 0.99 (0.04)  & 0.07 (0.14)  && 0.002 (0.003)  & 0.99 (0.02)  & 0.08 (0.15)  && 0.001 (0.003)  & 1.00 (0.01)  & 0.07 (0.14)  \\ 
  & GGL  &  0.114  &  0.000 && 0.169 (0.006)  & 0.10 (0.01)  & 0.81 (0.04)  && 0.122 (0.006)  & 0.13 (0.01)  & 0.81 (0.04)  && 0.095 (0.005)  & 0.18 (0.01)  & 0.85 (0.03)  && 0.074 (0.004)  & 0.23 (0.02)  & 0.85 (0.04)  \\ 
  & SSJGL & - & -  && 0.007 (0.001)  & 0.98 (0.03)  & 0.32 (0.03)  && 0.007 (0.001)  & 0.90 (0.04)  & 0.30 (0.03)  && 0.007 (0.001)  & 0.89 (0.04)  & 0.29 (0.03)  && 0.007 (0.001)  & 0.86 (0.04)  & 0.28 (0.03)  \\ 
  & stabJGL  &  0.166  &  0.025 && 0.017 (0.003)  & 0.68 (0.09)  & 0.55 (0.04)  && 0.012 (0.001)  & 0.87 (0.06)  & 0.50 (0.04)  && 0.011 (0.001)  & 0.92 (0.04)  & 0.52 (0.05)  && 0.010 (0.001)  & 0.95 (0.03)  & 0.49 (0.05)  \\ 
 \midrule 
40 $\%$ & Glasso &  0.202  & -  && 0.026 (0.007)  & 0.41 (0.08)  & 0.50 (0.06)  && 0.017 (0.005)  & 0.59 (0.11)  & 0.47 (0.08)  && 0.015 (0.004)  & 0.71 (0.11)  & 0.52 (0.07)  && 0.014 (0.003)  & 0.78 (0.09)  & 0.52 (0.08)  \\ 
  & FGL  &  0.114  &  0.003 && 0.146 (0.025)  & 0.11 (0.02)  & 0.80 (0.04)  && 0.104 (0.021)  & 0.16 (0.03)  & 0.79 (0.04)  && 0.078 (0.018)  & 0.23 (0.06)  & 0.83 (0.04)  && 0.059 (0.014)  & 0.30 (0.07)  & 0.83 (0.04)  \\ 
  & FGL (eBIC) &  0.489  &  0.000 && 0.000 (0.000)  & 1.00 (0.00)  & 0.00 (0.01)  && 0.000 (0.000)  & 1.00 (0.00)  & 0.00 (0.00)  && 0.000 (0.000)  & 1.00 (0.00)  & 0.00 (0.00)  && 0.000 (0.000)  & 1.00 (0.00)  & 0.00 (0.00)  \\ 
  & GGL  &  0.114  &  0.000 && 0.169 (0.005)  & 0.10 (0.00)  & 0.81 (0.04)  && 0.123 (0.005)  & 0.13 (0.01)  & 0.81 (0.04)  && 0.093 (0.005)  & 0.18 (0.01)  & 0.84 (0.03)  && 0.071 (0.004)  & 0.24 (0.01)  & 0.85 (0.03)  \\ 
  & SSJGL & - & -  && 0.004 (0.001)  & 0.93 (0.05)  & 0.19 (0.03)  && 0.004 (0.001)  & 0.69 (0.07)  & 0.14 (0.03)  && 0.004 (0.001)  & 0.77 (0.08)  & 0.15 (0.03)  && 0.004 (0.001)  & 0.81 (0.07)  & 0.16 (0.03)  \\ 
  & stabJGL  &  0.166  &  0.025 && 0.017 (0.004)  & 0.63 (0.10)  & 0.50 (0.05)  && 0.011 (0.002)  & 0.83 (0.07)  & 0.43 (0.05)  && 0.010 (0.001)  & 0.91 (0.04)  & 0.44 (0.05)  && 0.009 (0.001)  & 0.96 (0.03)  & 0.41 (0.05)  \\ 
 \midrule
20 $\%$ & Glasso &  0.203  & -  && 0.026 (0.007)  & 0.41 (0.08)  & 0.50 (0.06)  && 0.018 (0.004)  & 0.58 (0.09)  & 0.51 (0.06)  && 0.016 (0.003)  & 0.67 (0.07)  & 0.54 (0.06)  && 0.015 (0.003)  & 0.76 (0.09)  & 0.54 (0.07)  \\ 
  & FGL  &  0.114  &  0.002 && 0.154 (0.022)  & 0.11 (0.02)  & 0.80 (0.04)  && 0.113 (0.019)  & 0.15 (0.03)  & 0.82 (0.04)  && 0.085 (0.015)  & 0.21 (0.04)  & 0.84 (0.03)  && 0.065 (0.013)  & 0.27 (0.05)  & 0.85 (0.05)  \\ 
  & FGL (eBIC)  &  0.450  &  0.003 && 0.002 (0.004)  & 0.95 (0.11)  & 0.06 (0.14)  && 0.001 (0.003)  & 0.98 (0.04)  & 0.06 (0.14)  && 0.001 (0.003)  & 0.99 (0.02)  & 0.06 (0.14)  && 0.001 (0.003)  & 1.00 (0.01)  & 0.06 (0.13)  \\ 
  & GGL  &  0.114  &  0.000 && 0.169 (0.005)  & 0.10 (0.00)  & 0.81 (0.04)  && 0.125 (0.004)  & 0.13 (0.01)  & 0.84 (0.04)  && 0.095 (0.004)  & 0.18 (0.01)  & 0.86 (0.03)  && 0.074 (0.004)  & 0.23 (0.02)  & 0.86 (0.04)  \\ 
  & SSJGL & - & -  && 0.004 (0.000)  & 0.79 (0.07)  & 0.15 (0.02)  && 0.004 (0.000)  & 0.68 (0.07)  & 0.13 (0.02)  && 0.004 (0.001)  & 0.79 (0.06)  & 0.15 (0.02)  && 0.004 (0.001)  & 0.82 (0.06)  & 0.16 (0.02)  \\ 
  & stabJGL  &  0.166  &  0.023 && 0.017 (0.004)  & 0.59 (0.09)  & 0.48 (0.05)  && 0.012 (0.002)  & 0.78 (0.08)  & 0.46 (0.05)  && 0.010 (0.001)  & 0.89 (0.06)  & 0.46 (0.04)  && 0.009 (0.001)  & 0.96 (0.04)  & 0.42 (0.04)  \\ 
 \midrule
0 $\%$ & Glasso &  0.207  & -  && 0.027 (0.008)  & 0.40 (0.08)  & 0.51 (0.07)  && 0.019 (0.005)  & 0.70 (0.10)  & 0.65 (0.08)  && 0.018 (0.004)  & 0.80 (0.09)  & 0.70 (0.08)  && 0.015 (0.003)  & 0.85 (0.07)  & 0.63 (0.07)  \\ 
  & FGL  &  0.114  &  0.000 && 0.170 (0.005)  & 0.10 (0.00)  & 0.81 (0.04)  && 0.122 (0.004)  & 0.16 (0.01)  & 0.94 (0.02)  && 0.092 (0.004)  & 0.21 (0.01)  & 0.96 (0.02)  && 0.072 (0.003)  & 0.26 (0.01)  & 0.95 (0.02)  \\ 
  & FGL (eBIC) &  0.380  &  0.011 && 0.004 (0.005)  & 0.91 (0.13)  & 0.14 (0.16)  && 0.003 (0.004)  & 0.98 (0.04)  & 0.15 (0.18)  && 0.003 (0.004)  & 0.99 (0.04)  & 0.15 (0.19)  && 0.002 (0.003)  & 0.99 (0.02)  & 0.12 (0.15)  \\ 
  & GGL  &  0.114  &  0.000 && 0.170 (0.005)  & 0.10 (0.00)  & 0.81 (0.04)  && 0.122 (0.004)  & 0.16 (0.01)  & 0.94 (0.02)  && 0.092 (0.004)  & 0.21 (0.01)  & 0.96 (0.02)  && 0.072 (0.003)  & 0.26 (0.01)  & 0.95 (0.02)  \\ 
  & SSJGL & - & -  && 0.003 (0.000)  & 0.59 (0.10)  & 0.07 (0.02)  && 0.002 (0.000)  & 0.63 (0.10)  & 0.08 (0.02)  && 0.002 (0.000)  & 0.48 (0.09)  & 0.06 (0.01)  && 0.003 (0.000)  & 0.69 (0.10)  & 0.09 (0.02)  \\ 
  & stabJGL  &  0.166  &  0.015 && 0.026 (0.006)  & 0.42 (0.07)  & 0.52 (0.06)  && 0.018 (0.004)  & 0.73 (0.08)  & 0.65 (0.07)  && 0.016 (0.003)  & 0.85 (0.06)  & 0.66 (0.07)  && 0.013 (0.002)  & 0.89 (0.05)  & 0.57 (0.08)  \\
  \bottomrule
\end{tabular}}
\end{sidewaystable}

\begin{table}
\caption{Performance of the different graph reconstruction methods in simulations, reconstructing graphs with $p=100$ nodes from $K=2$ classes with various similarity of the true graph structures. The methods included are the graphical lasso (Glasso), the fused joint graphical lasso tuned by the AIC (FGL) and by the extended BIC (eBIC), the group joint graphical lasso (GGL), the Bayesian spike-and-slab joint graphical lasso (SSJGL) and stabJGL. The similarity (percentage of edges that are in common) of the graphs is shown. The results are averaged over $N=100$ simulations and shows the sparsity, precision, and recall of each of the $K=2$ estimated graphs. The corresponding standard deviations are shown as well. The graphs are reconstructed from $n_1=100$ and $n_2=150$ observations. All graphs have sparsity $0.02$. 
The average selected values of the penalty parameters $\lambda_1$ and $\lambda_2$ for the relevant methods is shown as well.  \label{table:simulationC}}
\renewcommand{\arraystretch}{1.3}
\resizebox{\textwidth}{!}{%
\begin{tabular}{r l r r l @{\hskip 0.1cm}l l l l@{\hskip 0.1cm}l l l}
\toprule
&&&&& \multicolumn{3}{@{}c@{}}{$n_1=100$}&&\multicolumn{3}{@{}c@{}}{$n_2=150$}\\
\cline{6-8}\cline{10-12} %
Similarity & Method & $\lambda_1$& $\lambda_2$ &&Sparsity& Precision & Recall &&Sparsity& Precision & Recall\\
\midrule
100 $\%$ & Glasso &  0.241  & -  && 0.021 (0.009)  & 0.38 (0.10)  & 0.36 (0.08)  && 0.024 (0.006)  & 0.43 (0.08)  & 0.50 (0.05)  \\ 
  & FGL  &  0.168  &  0.022 && 0.079 (0.023)  & 0.18 (0.06)  & 0.64 (0.05)  && 0.045 (0.013)  & 0.31 (0.09)  & 0.65 (0.05)  \\ 
  & FGL (eBIC) & 0.512  &  0.001 && 0.000 (0.000)  & -  & -  && 0.000 (0.000)  & - & - \\
  & GGL  &  0.167  &  0.014 && 0.088 (0.019)  & 0.15 (0.04)  & 0.61 (0.05)  && 0.049 (0.013)  & 0.28 (0.08)  & 0.63 (0.05)  \\ 
  & SSJGL & - & -  && 0.005 (0.001)  & 0.97 (0.04)  & 0.25 (0.05)  && 0.005 (0.001)  & 0.97 (0.04)  & 0.25 (0.05)  \\ 
  & stabJGL  &  0.218  &  0.095 && 0.014 (0.002)  & 0.69 (0.06)  & 0.47 (0.04)  && 0.012 (0.001)  & 0.78 (0.06)  & 0.47 (0.03)  \\ 
\midrule
80 $\%$ & Glasso &  0.238  & -  && 0.021 (0.009)  & 0.39 (0.11)  & 0.36 (0.07)  && 0.024 (0.008)  & 0.40 (0.10)  & 0.45 (0.07)  \\ 
  & FGL  &  0.170  &  0.017 && 0.082 (0.023)  & 0.16 (0.06)  & 0.62 (0.06)  && 0.042 (0.012)  & 0.31 (0.11)  & 0.58 (0.07)  \\ 
  & FGL (eBIC) & 0.504  &  0.002 && 0.000 (0.000)  & -  & -  && 0.000 (0.000)  & - & - \\
  & GGL  &  0.169  &  0.014 && 0.084 (0.020)  & 0.15 (0.04)  & 0.60 (0.06)  && 0.042 (0.012)  & 0.29 (0.08)  & 0.56 (0.07)  \\ 
  & SSJGL & - & -  && 0.004 (0.001)  & 0.97 (0.04)  & 0.19 (0.03)  && 0.004 (0.001)  & 0.97 (0.04)  & 0.19 (0.03)  \\ 
  & stabJGL  &  0.218  &  0.093 && 0.012 (0.002)  & 0.69 (0.07)  & 0.40 (0.04)  && 0.009 (0.001)  & 0.82 (0.05)  & 0.37 (0.04)  \\ 
\midrule
60 $\%$ & Glasso &  0.236  & -  && 0.022 (0.008)  & 0.36 (0.08)  & 0.36 (0.08)  && 0.026 (0.007)  & 0.39 (0.07)  & 0.48 (0.07)  \\ 
  & FGL  &  0.168  &  0.013 && 0.087 (0.019)  & 0.15 (0.04)  & 0.61 (0.05)  && 0.048 (0.011)  & 0.26 (0.06)  & 0.61 (0.06)  \\ 
  & FGL (eBIC) & 0.510  &  0.001 && 0.000 (0.000)  & -  & -  && 0.000 (0.000)  & - & - \\
  & GGL  &  0.167  &  0.012 && 0.089 (0.017)  & 0.14 (0.03)  & 0.61 (0.05)  && 0.049 (0.010)  & 0.26 (0.05)  & 0.60 (0.06)  \\ 
  & SSJGL & - & -  && 0.004 (0.001)  & 0.96 (0.05)  & 0.19 (0.03)  && 0.004 (0.001)  & 0.96 (0.04)  & 0.19 (0.03)  \\ 
  & stabJGL  &  0.218  &  0.092 && 0.012 (0.002)  & 0.66 (0.07)  & 0.39 (0.04)  && 0.010 (0.001)  & 0.78 (0.07)  & 0.38 (0.03)  \\ 
 \midrule
40 $\%$ & Glasso &  0.239  & -  && 0.021 (0.009)  & 0.38 (0.10)  & 0.36 (0.08)  && 0.025 (0.007)  & 0.41 (0.08)  & 0.50 (0.06)  \\ 
  & FGL  &  0.168  &  0.013 && 0.089 (0.020)  & 0.15 (0.04)  & 0.61 (0.06)  && 0.049 (0.012)  & 0.27 (0.07)  & 0.62 (0.06)  \\ 
  & FGL (eBIC) & 0.519  &  0.001 && 0.000 (0.000)  & -  & -  && 0.000 (0.000)  & - & - \\
  & GGL  &  0.167  &  0.012 && 0.089 (0.019)  & 0.14 (0.03)  & 0.60 (0.05)  && 0.049 (0.012)  & 0.26 (0.06)  & 0.61 (0.07)  \\ 
  & SSJGL & - & -  && 0.004 (0.001)  & 0.94 (0.05)  & 0.17 (0.03)  && 0.004 (0.001)  & 0.96 (0.05)  & 0.17 (0.03)  \\ 
  & stabJGL  &  0.218  &  0.092 && 0.011 (0.002)  & 0.66 (0.07)  & 0.36 (0.04)  && 0.009 (0.001)  & 0.79 (0.05)  & 0.35 (0.04)  \\ 
\midrule
20 $\%$ & Glasso &  0.239  & -  && 0.021 (0.009)  & 0.38 (0.11)  & 0.36 (0.08)  && 0.025 (0.007)  & 0.41 (0.09)  & 0.49 (0.06)  \\ 
  & FGL  &  0.168  &  0.009 && 0.094 (0.016)  & 0.13 (0.03)  & 0.60 (0.06)  && 0.051 (0.010)  & 0.25 (0.05)  & 0.62 (0.06)  \\ 
  & FGL (eBIC) & 0.508  &  0.002 && 0.000 (0.000)  & -  & -  && 0.000 (0.000)  & - & - \\
  & GGL  &  0.167  &  0.012 && 0.091 (0.018)  & 0.14 (0.03)  & 0.59 (0.06)  && 0.049 (0.012)  & 0.26 (0.06)  & 0.61 (0.06)  \\ 
  & SSJGL & - & -  && 0.003 (0.001)  & 0.88 (0.09)  & 0.12 (0.02)  && 0.003 (0.001)  & 0.93 (0.07)  & 0.12 (0.02)  \\ 
  & stabJGL  &  0.218  &  0.088 && 0.011 (0.003)  & 0.60 (0.07)  & 0.33 (0.05)  && 0.008 (0.001)  & 0.76 (0.08)  & 0.30 (0.05)  \\ 
  \midrule
0 $\%$ & Glasso &  0.238  & -  && 0.022 (0.009)  & 0.38 (0.10)  & 0.37 (0.08)  && 0.025 (0.007)  & 0.43 (0.08)  & 0.51 (0.07)  \\ 
  & FGL  &  0.167  &  0.006 && 0.098 (0.015)  & 0.13 (0.02)  & 0.62 (0.06)  && 0.053 (0.009)  & 0.26 (0.04)  & 0.66 (0.06)  \\ 
  & FGL (eBIC) & 0.520  &  0.002 && 0.000 (0.000)  & -  & -  && 0.000 (0.000)  & - & - \\
  & GGL  &  0.167  &  0.008 && 0.095 (0.018)  & 0.13 (0.03)  & 0.61 (0.06)  && 0.051 (0.011)  & 0.27 (0.05)  & 0.65 (0.07)  \\ 
  & SSJGL & - & -  && 0.002 (0.000)  & 0.79 (0.11)  & 0.09 (0.02)  && 0.002 (0.000)  & 0.90 (0.08)  & 0.10 (0.02)  \\ 
  & stabJGL  &  0.218  &  0.088 && 0.010 (0.002)  & 0.58 (0.07)  & 0.29 (0.05)  && 0.007 (0.001)  & 0.76 (0.08)  & 0.26 (0.04)  \\ 
\bottomrule
\end{tabular}}
\end{table}

\begin{table}
\caption{Performance of the different graph reconstruction methods in simulations, reconstructing graphs with $p=300$ nodes from $K=2$ classes with various similarity of the true graph structures. The methods included are the graphical lasso (Glasso), the fused joint graphical lasso tuned by the AIC (FGL) and by the extended BIC (eBIC), the group joint graphical lasso (GGL) and stabJGL. The similarity (percentage of edges that are in common) of the graphs is shown. The results are averaged over $N=100$ simulations and shows the sparsity, precision, and recall of each of the $K=2$ estimated graphs. The corresponding standard deviations are shown as well. The graphs are reconstructed from $n_1=150$ and $n_2=200$ observations. All graphs have sparsity $0.007$. 
The average selected values of the penalty parameters $\lambda_1$ and $\lambda_2$ for the relevant methods is shown as well.  \label{table:simulationD}}
\renewcommand{\arraystretch}{1.3}
\resizebox{\textwidth}{!}{%
\begin{tabular}{r l r r l @{\hskip 0.1cm}l l l l@{\hskip 0.1cm}l l l}
\toprule
&&&&& \multicolumn{3}{@{}c@{}}{$n_1=150$}&&\multicolumn{3}{@{}c@{}}{$n_2=200$}\\
\cline{6-8}\cline{10-12} %
Similarity & Method & $\lambda_1$& $\lambda_2$ &&Sparsity& Precision & Recall &&Sparsity& Precision & Recall\\
\midrule
100 $\%$ & Glasso &  0.195  & -  && 0.009 (0.003)  & 0.26 (0.07)  & 0.30 (0.04)  && 0.006 (0.002)  & 0.39 (0.09)  & 0.31 (0.04)  \\ 
  & FGL  &  0.166  &  0.022 && 0.012 (0.005)  & 0.25 (0.11)  & 0.39 (0.03)  && 0.007 (0.003)  & 0.41 (0.14)  & 0.37 (0.03)  \\ 
  & FGL (eBIC) &  0.453  &  0.001 && 0.000 (0.000)  & -  & - && 0.000 (0.000)  & -  & -  \\ 
  & GGL  &  0.166  &  0.004 && 0.020 (0.003)  & 0.14 (0.02)  & 0.40 (0.03)  && 0.011 (0.002)  & 0.24 (0.04)  & 0.39 (0.03)  \\ 
  & stabJGL  &  0.166  &  0.098 && 0.003 (0.000)  & 0.69 (0.03)  & 0.35 (0.02)  && 0.003 (0.000)  & 0.72 (0.04)  & 0.35 (0.02)  \\ 
\midrule
80 $\%$ & Glasso &  0.195  & -  && 0.008 (0.003)  & 0.27 (0.07)  & 0.30 (0.04)  && 0.006 (0.003)  & 0.39 (0.14)  & 0.30 (0.06)  \\ 
  & FGL  &  0.166  &  0.012 && 0.016 (0.005)  & 0.18 (0.06)  & 0.39 (0.03)  && 0.008 (0.003)  & 0.32 (0.10)  & 0.37 (0.03)  \\ 
  & FGL (eBIC) &  0.442  &  0.002 && 0.000 (0.000)  & -  & -  && 0.000 (0.000)  & -  & -  \\ 
  & GGL  &  0.166  &  0.005 && 0.019 (0.004)  & 0.14 (0.03)  & 0.40 (0.03)  && 0.010 (0.002)  & 0.26 (0.05)  & 0.37 (0.03)  \\ 
  & stabJGL  &  0.166  &  0.093 && 0.003 (0.000)  & 0.69 (0.04)  & 0.32 (0.02)  && 0.003 (0.000)  & 0.72 (0.04)  & 0.30 (0.02)  \\ 
\midrule
60 $\%$ & Glasso &  0.195  & -  && 0.008 (0.003)  & 0.26 (0.07)  & 0.30 (0.04)  && 0.006 (0.002)  & 0.38 (0.11)  & 0.31 (0.05)  \\ 
  & FGL  &  0.166  &  0.010 && 0.017 (0.005)  & 0.17 (0.05)  & 0.40 (0.03)  && 0.009 (0.003)  & 0.30 (0.08)  & 0.37 (0.03)  \\ 
  & FGL (eBIC) &   0.449  &  0.002 && 0.000 (0.000)  & - & -  && 0.000 (0.000)  & -  & -  \\ 
  & GGL  &  0.166  &  0.005 && 0.020 (0.004)  & 0.14 (0.02)  & 0.40 (0.03)  && 0.011 (0.002)  & 0.25 (0.05)  & 0.38 (0.03)  \\ 
  & stabJGL  &  0.166  &  0.091 && 0.003 (0.000)  & 0.65 (0.04)  & 0.29 (0.02)  && 0.003 (0.000)  & 0.70 (0.04)  & 0.28 (0.02)  \\ 
\midrule
40 $\%$ & Glasso &  0.196  & -  && 0.009 (0.003)  & 0.26 (0.08)  & 0.30 (0.05)  && 0.006 (0.003)  & 0.41 (0.13)  & 0.29 (0.06)  \\ 
  & FGL  &  0.166  &  0.007 && 0.018 (0.004)  & 0.15 (0.04)  & 0.40 (0.03)  && 0.009 (0.002)  & 0.27 (0.07)  & 0.36 (0.03)  \\ 
  & FGL (eBIC) & 0.446  &  0.002 && 0.000 (0.000)  & - & -  && 0.000 (0.000)  & -  & - \\ 
  & GGL  &  0.166  &  0.004 && 0.020 (0.004)  & 0.14 (0.03)  & 0.40 (0.03)  && 0.010 (0.002)  & 0.25 (0.05)  & 0.37 (0.03)  \\ 
  & stabJGL  &  0.166  &  0.091 && 0.003 (0.000)  & 0.64 (0.05)  & 0.25 (0.02)  && 0.002 (0.000)  & 0.68 (0.04)  & 0.23 (0.02)  \\ 
\midrule
20 $\%$ & Glasso &  0.195  & -  && 0.008 (0.003)  & 0.28 (0.08)  & 0.30 (0.04)  && 0.006 (0.003)  & 0.38 (0.13)  & 0.30 (0.06)  \\ 
  & FGL  &  0.166  &  0.006 && 0.019 (0.004)  & 0.14 (0.03)  & 0.39 (0.03)  && 0.010 (0.002)  & 0.26 (0.06)  & 0.37 (0.04)  \\ 
  & FGL (eBIC) &  0.450  &  0.001 && 0.000 (0.000)  & -  & - && 0.000 (0.000)  & - & -  \\ 
  & GGL  &  0.166  &  0.004 && 0.020 (0.003)  & 0.14 (0.02)  & 0.40 (0.03)  && 0.010 (0.002)  & 0.25 (0.04)  & 0.37 (0.03)  \\ 
  & stabJGL  &  0.166  &  0.091 && 0.002 (0.000)  & 0.60 (0.05)  & 0.22 (0.02)  && 0.002 (0.000)  & 0.63 (0.05)  & 0.19 (0.02)  \\ 
\midrule
0 $\%$ & Glasso &  0.198  & -  && 0.008 (0.003)  & 0.27 (0.07)  & 0.30 (0.04)  && 0.005 (0.002)  & 0.50 (0.12)  & 0.37 (0.07)  \\ 
  & FGL  &  0.166  &  0.000 && 0.022 (0.001)  & 0.13 (0.01)  & 0.42 (0.02)  && 0.012 (0.001)  & 0.28 (0.02)  & 0.51 (0.03)  \\ 
  & FGL (eBIC) &  0.462  &  0.002 && 0.000 (0.000)  & -  & -  && 0.000 (0.000)  & -  & - \\ 
  & GGL  &  0.166  &  0.001 && 0.022 (0.002)  & 0.13 (0.01)  & 0.41 (0.02)  && 0.012 (0.001)  & 0.28 (0.02)  & 0.51 (0.03)  \\ 
  & stabJGL  &  0.166  &  0.077 && 0.003 (0.001)  & 0.48 (0.05)  & 0.20 (0.03)  && 0.002 (0.000)  & 0.62 (0.04)  & 0.21 (0.03)  \\ 
\bottomrule
\end{tabular}}
\end{table}

\section{Choice of variability threshold}

\begin{figure*}[t]
    \centering
    \includegraphics[width=0.8\textwidth]{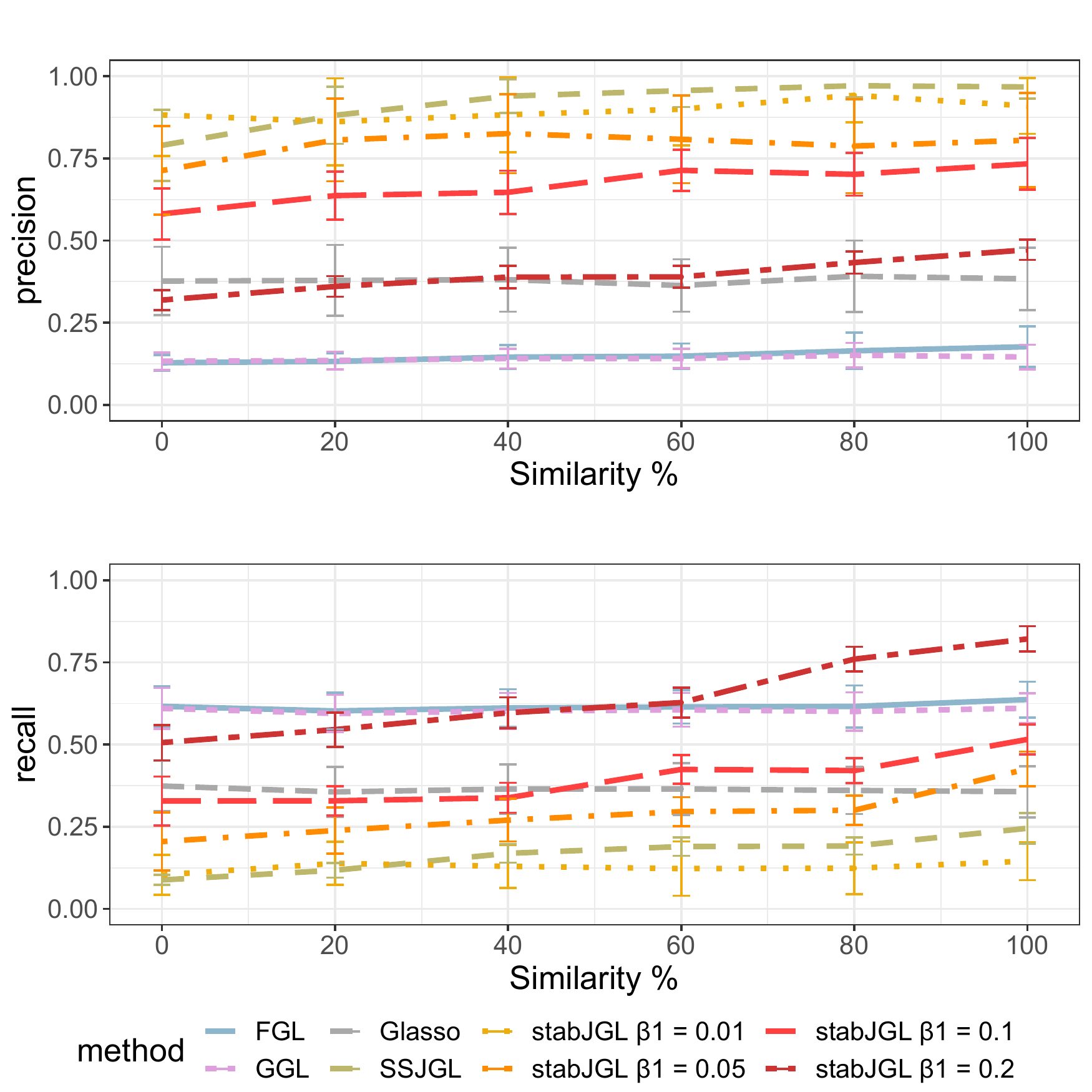}
    \caption{Performance of stabJGL for different values of the variability threshold $\beta_1$ on simulated data, compared to other graph reconstruction methods. The methods are used to estimate graphs with $p=100$ nodes from $K=2$ networks, both of sparsity $0.02$, with various similarity of the true graph structures. The performance of stabJGL is compared to that of the graphical lasso (Glasso), the fused joint graphical lasso tuned by the AIC (FGL), the group joint graphical lasso (GGL) and the Bayesian spike-and-slab joint graphical lasso (SSJGL). The similarity (percentage of edges that are in common) of the graphs is shown. The results are averaged over $N=100$ simulations and shows the precision and recall of each of the $K=2$ estimated graphs. Standard deviation bars are shown for all methods. The graphs are reconstructed from $n_1=100$ and $n_2=150$ observations.}
    \label{fig:simstudy_thresh_full}
\end{figure*}

Figure \ref{fig:simstudy_thresh_full} compares the performance of stabJGL for different values of the variability threshold $\beta_1$ to the the graphical lasso (Glasso), the fused joint graphical lasso (FGL), the group joint graphical lasso (GGL) and the Bayesian spike-and-slab joint graphical lasso (SSJGL). The results for FGL tuned with eBIC are not shown as it selected an empty graph in all settings. The settings considered have $K=2$ networks with $p=100$ nodes of various similarity. As in the setting considered in the main manuscript, we find that by varying the variability threshold $\beta_1$ we can obtain at least as high precision and/or recall as the other methods at any level of similarity.

\section{Additional Pan-Cancer analysis results}

\subsection{Degree distributions}

\begin{figure}[t]
    \centering
    \includegraphics[width=1\textwidth]{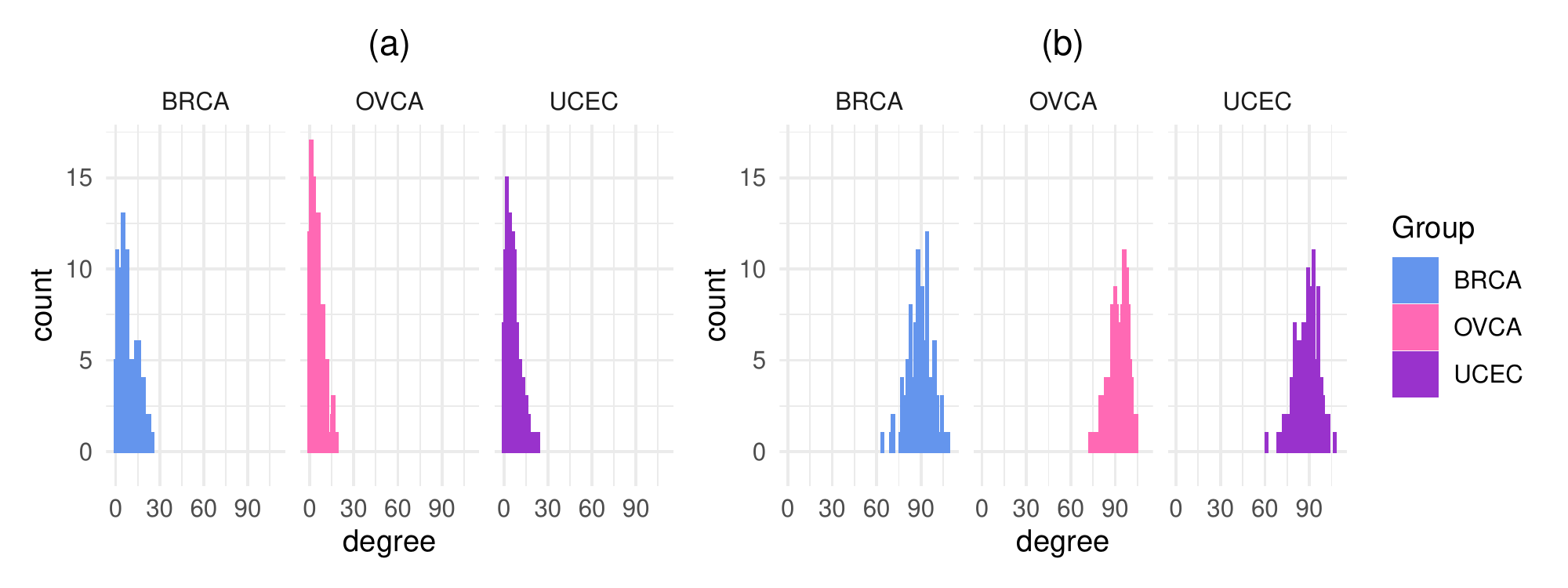}
    \caption{Histogram of the node degrees of the proteomic network of each tumor type, for the (a) stabJGL and (b) FGL networks.}
    \label{fig:PanCan_degreehist}
\end{figure}

Table \ref{fig:PanCan_degreehist} shows the degree distribution of the proteomic networks identified by stabJGL and FGL. While the stabJGL networks all have degree distributions that follow clear power-law distributions, in line with biological expectations, the FGL networks have degree distributions that strongly contradict a power law with most nodes having node degree $>60$.

\subsection{Top hubs}

Table \ref{table:PanCan_topgenes} shows the node degree of the proteins with degree larger than the $90^{\text{th}}$ percentile in the respective stabJGL networks of the different tumor types. The same table for the FGL networks is shown in Table \ref{table:PanCan_topgenes}.

\begin{table}
	\centering
	\renewcommand{\arraystretch}{1.5}
	\caption{The genes with node degree larger than the $90^{\text{th}}$ percentile in the respective stabJGL networks of the different tumor types. The genes that have node degree in the upper $10\%$ in all three tumor types are marked in bold. The genes that only have node degree in the upper $10\%$ in one tumor type are marked in red.}
	\hspace*{-1cm}
	\resizebox{1.1\textwidth}{!}{
	\begin{tabular}{l l  r r @{\hskip 0.8cm}l l r r @{\hskip 0.8cm}l l r }
        \toprule
	    \multicolumn{3}{c}{BRCA}&&\multicolumn{3}{c}{UCEC}&&\multicolumn{3}{c}{OVCA} \\
		\cmidrule(lr){1-3} \cmidrule(lr){5-7} \cmidrule(lr){9-11}
		Protein & Gene & Degree && Protein &Gene & Degree && Protein &Gene & Degree \\
		\hline
       \color{red}{mTOR} & \emph{MTOR} & 25 && \textbf{14-3-3-epsilon} & \emph{YWHAE} & 23 && \color{red}{Bak} & \emph{BAK1} & 18 \\ 
  \textbf{14-3-3-epsilon} & \emph{YWHAE} & 24 && \color{red}{CD31} & \emph{PECAM1} & 20 && MRE11 & \emph{MRE11A} & 17 \\ 
 \color{red}{EGFR} & \emph{EGFR} & 23 && MRE11 & \emph{MRE11A} & 19 && \textbf{EGFR-pY1173} & \emph{EGFR} & 16 \\ 
  Chk1 & \emph{CHEK1} & 23 && GSK-3-alpha-beta-pS21S9 & \emph{GSK3A} & 17 && \textbf{Bid} & \emph{BID} & 16 \\ 
  \color{red}{Tuberin} & \emph{TSC2} & 22 && \textbf{EGFR-pY1173} & \emph{EGFR} & 17 && \textbf{14-3-3-epsilon} & \emph{YWHAE} & 16 \\ 
  \textbf{EGFR-pY1173} & \emph{EGFR} & 22 && \color{red}{p38-pT180-Y182} & \emph{MAPK14} & 16 && \textbf{Stathmin} & \emph{STMN1} & 15 \\ 
  \textbf{Stathmin} & \emph{STMN1} & 21 && \textbf{Stathmin} & \emph{STMN1} & 15 && MAPK-pT202-Y204  & \emph{MAPK1} & 15 \\ 
 \color{red}{c-KIT} & \emph{KIT} & 21 && \color{red}{LKB1} & \emph{STK11} & 15 && GSK-3-alpha-beta-pS21S9  & \emph{GSK3A} & 13 \\ 
 \color{red}{Ku80} & \emph{XRCC5} & 20 && \textbf{Bid} & \emph{BID} & 15 && \color{red}{SMAD4} & \emph{SMAD4} & 12 \\ 
  \color{red}{S6} & \emph{RPS6} & 19 && MIG6 & \emph{ERRFI1} & 14 && MIG6 & \emph{ERRFI1} & 12 \\ 
 \color{red}{Hsp70} & \emph{HSPA1A} & 19 && MAPK-pT202-Y204 & \emph{MAPK1} & 14 && \color{red}{Cyclin D1} & \emph{CCND1} & 12 \\ 
  \color{red}{Collagen VI} & \emph{COL6A1} & 19 && c-Met-pY1235 & \emph{MET} & 14 && Chk1 & \emph{CHEK1} & 12 \\ 
 \textbf{Bid} & \emph{BID} & 19 &&  & & && c-Met-pY1235  & \emph{MET} & 12 \\ 
        \toprule
	\end{tabular}
	}
	\label{table:PanCan_topgenes}
\end{table} 

\begin{table}
	\centering
	\renewcommand{\arraystretch}{1.5}
	\caption{The genes with node degree larger than the $90^{\text{th}}$ percentile in the respective FGL networks of the different tumor types. The genes that have node degree in the upper $10\%$ in all three tumor types are marked in bold. The genes that only have node degree in the upper $10\%$ in one tumor type are marked in red.}
	\hspace*{-1cm}
	\resizebox{1.1\textwidth}{!}{
	\begin{tabular}{l l  r r @{\hskip 0.8cm}l l r r @{\hskip 0.8cm}l l r }
        \toprule
	    \multicolumn{3}{c}{BRCA}&&\multicolumn{3}{c}{UCEC}&&\multicolumn{3}{c}{OVCA} \\
		\cmidrule(lr){1-3} \cmidrule(lr){5-7} \cmidrule(lr){9-11}
		Protein & Gene & Degree && Protein &Gene & Degree && Protein &Gene & Degree \\
		\hline
        \color{red}{NF-kB-p65-pS536} & \emph{NFKB1} & 108 && \color{red}{Src} & \emph{SRC} & 107 && \color{red}{SYK} & \emph{SYK} & 104 \\ 
  \color{red}{AR} & \emph{AR} & 106 && \color{red}{PEA15} & \emph{PEA-15} & 103 && \color{red}{p70-S6K} & \emph{RPS6KB1} & 104 \\ 
  \color{red}{PAI-1} & \emph{SERPINE1} & 104 && \color{red}{JNK2} & \emph{MAPK9} & 103 && \color{red}{mTOR-pS2448} & \emph{MTOR} & 103 \\ 
  \color{red}{XRCC1} & \emph{XRCC1} & 104 && HER3 & \emph{ERBB3} & 102 && \color{red}{ER-alpha-pS118} & \emph{ESR1} & 102 \\ 
  \color{red}{Cyclin E1} & \emph{CCNE1} & 104 && \color{red}{c-Raf} & \emph{RAF1} & 102 && \color{red}{ATM} & \emph{ATM} & 102 \\ 
  \color{red}{S6} & \emph{RPS6} & 102 && \color{red}{VEGFR2} & \emph{KDR} & 100 && \color{red}{STAT5-alpha} & \emph{STAT5A} & 101 \\ 
  \color{red}{RAD50} & \emph{RAD50} & 101 && \color{red}{p53} & \emph{TP53} & 99 && \color{red}{PCNA} & \emph{PCNA} & 101 \\ 
  MEK1 & \emph{MAP2K1} & 101 && \color{red}{GSK-3-alpha-beta} & \emph{GSK3A} & 99 && MEK1 & \emph{MAP2K1} & 101 \\ 
 \color{red}{INPP4B} & \emph{INPP4B} & 101 && \color{red}{AMPK-pT172} & \emph{PRKAA1} & 99 && \color{red}{53BP1} & \emph{TP53BP1} & 101 \\ 
  \color{red}{MIG6} & \emph{ERRFI1} & 100 && \color{red}{Tuberin} & \emph{TSC2} & 98 &&  & & \\ 
HER3 & \emph{ERBB3} & 100 && \color{red}{p38-MAPK} & \emph{MAPK14} & 98 &&  & & \\ 
 & & && \color{red}{p27} & \emph{CDKN1B} & 98 &&  & & \\ 
 & & && \color{red}{IGFBP2} & \emph{IGFBP2} & 98 &&  & & \\ 
        \toprule
	\end{tabular}
	}
	\label{table:PanCan_topgenes_FGL}
\end{table}

\end{document}